%% file: Karczewski_n4449_paper1.tex
\documentclass[useAMS,usenatbib]{mn2e}

\usepackage{amssymb}
\usepackage{amsmath}
\usepackage{graphicx}
\usepackage{units}
\usepackage{multirow}
\usepackage{rotating}
\usepackage{hyperref}

\newcommand\farcsec{$.\!\!^{\rm s}$}
\newcommand\farcsecp{$.\!\!^{\rm \prime\prime}$}
\newcommand\farcmin{$.\!\!^{\rm \prime}$}

\def\one{\,{\sc i}}
\def\two{\,{\sc ii}}
\def\three{\,{\sc iii}}

\newcommand{\Swift}{\textit{Swift}}
\newcommand{\Spitzer}{\textit{Spitzer}}
\newcommand{\Herschel}{\textit{Herschel}}
\newcommand{\Planck}{\textit{Planck}}
\newcommand{\GALEX}{\textit{GALEX}}
\newcommand{\WISE}{\textit{WISE}}
\newcommand{\DGR}{$M_{\rm dust}$/$M_{\rm gas}$}
\newcommand{\Lstar}{$L_\star$}

\defcitealias{Karczewski2013_OBSERVATIONS}{Paper II}
\bibpunct[; ]{(}{)}{;}{a}{}{,}

\title[A study of NGC 4449 -- I. The SED]{A multiwavelength study of the Magellanic-type galaxy NGC~4449 -- I. Modelling the spectral energy distribution, the ionization structure and the star formation history}
\author[O.~\L. Karczewski et al.]
{O.~\L. Karczewski$^1$\thanks{E-mail: olk@star.ucl.ac.uk}, M.~J. Barlow$^1$, M.~J. Page$^2$, 
N.~P.~M. Kuin$^2$, I. Ferreras$^2$, \newauthor
M. Baes$^3$, G.~J. Bendo$^4$, A. Boselli$^5$, A. Cooray$^6$, D. Cormier$^7$, I. De~Looze$^3$, 
\newauthor
M. Galametz$^8$, F. Galliano$^7$, V. Lebouteiller$^7$, S.~C. Madden$^7$, M. Pohlen$^9$, 
\newauthor
A. R\'{e}my-Ruyer$^7$, M.~W.~L. Smith$^9$ and L. Spinoglio$^{10}$\\
$^1$Department of Physics and Astronomy, University College London, Gower Street, London, WC1E 6BT, UK\\
$^2$Mullard Space Science Laboratory, University College London, Holmbury St. Mary, Dorking, RH5 6NT, UK\\
$^3$Sterrenkundig Observatorium, Universiteit Gent, Krijgslaan 281-S9, 9000 Gent, Belgium\\
$^4$UK ALMA Regional Centre Node, Jodrell Bank Centre for Astrophysics, University of Manchester, Oxford Road,\\
$\;\,$Manchester, M13 9PL, UK\\
$^5$LAM, Universit\'{e} d'Aix-Marseille \& CNRS, UMR7326, 38 rue F. Joliot-Curie, 13388 Marseille, France\\
$^6$Department of Physics \& Astronomy, University of California, Irvine, CA 92697, USA\\
$^7$Laboratoire AIM, CEA/DSM--CNRS--Universit\'{e} Paris Diderot, IRFU/Service d'Astrophysique, CEA Saclay, \\ $\;\,$L'Orme des Merisiers, B\^{a}t. 709, 91191 Gif-sur-Yvette, France \\
$^8$Institute of Astronomy, University of Cambridge, Madingley Road, Cambridge, CB3 0HA, UK\\
$^9$School of Physics and Astronomy, Cardiff University, The Parade, Cardiff, CF24 3AA, UK\\
$^{10}$Istituto di Astrofisica e Planetologia Spaziali, INAF-IAPS, Via Fosso del Cavaliere 100, 00133 Roma, Italy\\
}

\begin{document}

\date{Accepted: Received:}

\pagerange{\pageref{firstpage}--\pageref{lastpage}} \pubyear{2002}

\maketitle

\label{firstpage}

\begin{abstract}

We present an integrated photometric spectral energy distribution (SED)
of the Magellanic-type galaxy NGC~4449 from the far-ultraviolet to the submillimetre,
including new observations acquired by the {\it Herschel} Space Observatory.
We include integrated UV photometry from the
\Swift\ Ultraviolet and Optical Telescope using
a measurement technique which is appropriate for extended sources
with coincidence loss.

In this paper, we examine the available multiwavelength data
to infer a range of ages, metallicities and star formation
rates for the underlying stellar populations, as well as the composition and the total
mass of dust in NGC~4449.
Our analysis of the global optical spectrum of NGC~4449 fitted
using the spectral fitting code STARLIGHT suggests
that the majority of stellar mass resides in old ($\gtrsim 1$~Gyr old)
and metal-poor ($Z/\textrm{Z}_\odot$$\,\,\sim\,\,$$0.2$) populations,
with the first onset of star formation activity deduced to have
taken place at an early epoch, approximately 12 Gyr~ago. A simple chemical evolution model,
suitable for a galaxy continuously forming stars,
suggests a ratio of carbon to silicate dust mass comparable to that of the
Large Magellanic Cloud over the inferred time-scales.

We present an iterative scheme, which
allows us to build an in-depth and
multicomponent representation of NGC~4449 `bottom-up',
taking advantage of the broad capabilities of the 
photoionization and radiative transfer code MOCASSIN
(MOnte CArlo SimulationS of Ionized Nebulae).
We fit the observed SED, the global ionization structure 
and the emission line intensities, and
infer a recent star formation rate of 0.4 $\textrm{M}_\odot\,\rm{yr}^{-1}$
and a total stellar mass of $\approx$$\;1\times10^9\;\textrm{M}_\odot$
emitting with a bolometric luminosity of 5.7$\;\times\,$10$^9\;\textrm{L}_\odot$.
Our fits yield a total dust mass of $2.9\pm0.5\times10^6\;\textrm{M}_\odot$ including
2 per~cent attributed to polycyclic aromatic hydrocarbons.
We deduce
a dust to gas mass ratio of 1/190 within the modelled
region. While we do not consider possible additional
contributions from even colder dust, we note that including the
extended H\one\ envelope and the molecular gas is likely to bring
the ratio down to as low as $\sim\;$1/800.

\end{abstract}


\section{Introduction}

The conditions in the interstellar medium (ISM) in a galaxy and the
environment-dependent processes governing the formation of stars and dust
can be studied on many physical scales.
Considering an entire galaxy, the interplay between 
the underlying stellar emission, the degree of radiation reprocessing taking
place in the ISM and the thermal or non-thermal emission due to
dynamical or evolutionary processes give rise to a
unique pattern of emission and absorption features in
the observed spectral energy distribution (SED).

The spectroscopic measurements alone provide a wealth
of information about the ionized (H\two) regions or
photo-dissociation regions (PDRs) in galaxies (e.g.
\citealt{Guseva2004}; \citealt{Vasta2010}).
Studies constrained
by broadband photometric measurements
can allow a better understanding of the
main components of the thermal continuum,
namely, the stellar emission and the emission due to interstellar dust.
Observed emission lines and diagnostic
line ratios provide important
further constraints \citep{Martinez-Galarza2011}.
Synthetic single stellar populations (SSPs)
can be used to characterise
star-forming regions within galaxies (e.g. \citealt{Groves2008},
\citealt{Ferreras2012}), or to decompose the observed SEDs of entire galaxies
(e.g. \citealt{Amorin2012}). The composition
of dust and the masses of dust species
can be inferred from observations by assuming
measured laboratory properties of dust (e.g.
\citealt{Galliano2008a} and references therein).
Together, these components can provide
an in-depth view of a galaxy (e.g. \citealt{Cormier2012}, \citealt{Hermelo2013}).

Numerical models reproducing
all the available photometric and spectroscopic measurements
across a wide range of wavelengths
can provide the most complete picture of the integrated
properties of a galaxy. We wish to
examine a multicomponent model treating
the stellar content, the gaseous phase and the dust
within a galaxy in a self-consistent way. Our initial goals
include finding a robust numerical scheme, which
could be applied to a sample of dwarf galaxies
in order to study the details of their
star formation histories and dust content.

In this paper, we describe the first multicomponent
observation-driven model
of a galaxy generated with the photoionization and radiative
transfer code 
MOCASSIN \citep{Ercolano2003,Ercolano2005,Ercolano2008},
which includes a simultaneous and self-consistent treatment of
the stellar component, the gaseous phase, its ionization
structure and multiple dust species. 
MOCASSIN is a fully three-dimensional code, whose
recent applications include modelling the extreme bipolar planetary nebula
NGC~6302 \citep{Wright2011} and studying dust emission
by supernova ejecta \citep{Wesson2010}. Applying MOCASSIN to simulate
entire galaxies is more resource-consuming since both
the gas phase and the dust phase need to be treated self-consistently
over relatively large physical scales. The small intrinsic size
of dwarf galaxies make them ideal candidates for such multiwavelength studies.
Indeed, dwarf galaxies provide small self-contained but dynamic
environments. As such, they are useful in developing our understanding of
mechanisms underlying star formation, including
chemical enrichment and feedback processes, especially
for low-metallicity objects resembling those formed at earlier
epochs in the evolution of the Universe. 

In a similar context, MOCASSIN was first used to
model Mrk~996 \citep{James2009}. In this paper we describe
a model of the well-known starburst galaxy NGC~4449, which is
a low-metallicity [$\log(\textrm{O}/\textrm{H})+12=8.23$, 1/3Z$_\odot$; \citealt{Engelbracht2008}] actively star-forming (SFR$\;\sim\;$0.5 $\textrm{M}_\odot\,\rm{yr}^{-1}$; \citealt{Hunter1998}, \citealt{Hill1998}) barred Magellanic-type irregular galaxy (\citealt{RC3}, \citealt{RC3_additions}) seen face-on. At a distance of 3.8 Mpc \citep{Annibali2008} its H\one\ envelope extends to 12.9 kpc (11\farcmin6), equal to approximately three Holmberg radii (\citealt{Swaters2002}, \citealt{Hunter1999}), and forms a pronounced system of streamers with a counter-rotating H\one\ core, which may be indicative of past mergers (\citealt{Hunter1999}, \citealt{Theis2001}). NGC~4449 is particularly suitable for multiwavelength studies because of its wide range of existing photometric data extending to the far-infrared (FIR). The new observations acquired with the {\it Spitzer} Space Telescope \citep{Spitzer} and the \textit{Herschel} Space Observatory \citep{Herschel} are essential in studying gas cooling processes and the properties of dust in the ISM of NGC~4449.

In the following sections we discuss in detail a model of NGC~4449
based on multiple observational constraints and including a
treatment of polycyclic aromatic hydrocarbons (PAHs).
Thus, in Section~\ref{sec:observations} we describe the
observations and data reduction techniques, and in
Section~\ref{sec:method} we describe the modelling method,
as well as the assumptions and the convergence criteria adopted.
The results and discussion are presented in Section~\ref{sec:results},
and the conclusions follow in Section~\ref{sec:conclusions}.
In a companion paper, \citet[hereafter, \citetalias{Karczewski2013_OBSERVATIONS}]{Karczewski2013_OBSERVATIONS} we
study the FIR cooling lines and discuss properties
of PDRs inferred from spatially-resolved spectroscopic observations of NGC 4449.

\section{Photometric Observations and Data Reduction}\label{sec:observations}

In constructing the observed SED of NGC 4449 we have used
data from previous studies (e.g, \citealt{Kennicutt1992}, \citealt{Engelbracht2008}, \citealt{Bendo2012b}, \citealt{Hunter1986}, \citealt{Bottner2003}) and from systematic surveys [Sloan
Digital Sky Survey (SDSS), \citealt{SDSS}, Two Micron All Sky
Survey (2MASS), \citealt{2MASS}, \WISE, \citealt{WISE} and \Planck, \citealt{Planck_ERCSC}]. We present archival data from \GALEX\ (\citealt{GALEX}; UV wavebands) and \Swift\ (\citealt{Swift}; UV and optical wavebands), as well as new FIR photometry from \textit{Herschel} \citep{Herschel}. A journal of observations is given in Table~\ref{tab:observations_table}.

In Table~\ref{tab:photometry_table} we list all global photometric
measurements used to construct the observed SED for NGC~4449. Existing
submillimetre SCUBA (Submillimetre Common-User Bolometer Array)
measurements at 450~$\mu$m and 850~$\mu$m \citep{Bottner2003}
were omitted from the list of photometric measurements due to their incomplete
field-of-view (FOV) and the availability of space borne \Herschel\ and \Planck\ observations
offering a similar wavelength coverage. Five illustrative false-colour images
of NGC~4449 over selected broadband ranges are presented in Fig.~\ref{fig:n4449_5imgs}.

\subsection{Ultraviolet observations with \GALEX\ and \Swift}

At UV wavelengths NGC~4449 has been observed by two recent space-borne
missions, \GALEX\ (in two UV bands) and the \Swift\ Ultraviolet/Optical
Telescope (UVOT; in three UV and three optical bands).
Standard aperture photometry using the in-flight calibration of
\citet{Morrissey2007} was performed on tile 5228 acquired as part of
\GALEX\ Nearby Galaxy Survey and accessible via General Release~6.
The overall uncertainties include photometric repeatability
measurements by \citet{Morrissey2007}.

\Swift/UVOT \citep{Swift_UVOT} offers four times the angular
resolution of \GALEX\ in three narrow UV bands. The telescope is
equipped with a photon-counting detector and was originally designed
to detect and observe gamma-ray bursts. As a consequence, standard
data reduction procedures cannot be applied straightforwardly to
observations of extended sources. In the Appendix we discuss this
problem in detail and present a method of obtaining global photometry,
which is appropriate for extended sources.

\subsection{Optical observations with SDSS}

In addition to the optical fluxes derived from the \Swift/UVOT
observations, SDSS observations in five optical
bands (\textit{u}, \textit{g}, \textit{r}, \textit{i} and \textit{z})
are available as part of Data Release~7. NGC~4449 is contained
entirely within one field, thus minimizing deblending or sky subtraction
issues which may arise in the case of extended sources. Standard aperture
photometry was performed using elliptical apertures and the counts were
converted into flux densities as described by \citet{West2010}.
The uncertainties in the derived fluxes are dominated by uncertainties
in sky determination and subtraction.

\input{./figures/figure.n4449.5images.tex}

\input{./tables/table.observations1.tex}

\input{./tables/table.photometry.tex}

\subsection{Far-infrared observations with \Herschel} 

NGC~4449 has been observed by \Herschel\ \citep{Herschel} as part of the Dwarf Galaxy Survey (DGS; \citealt{HerschelDGS}) with its two imaging photometers: the Photodetector Array Camera and Spectrometer (PACS; \citealt{Herschel_PACS}) at 70~$\mu$m, 100~$\mu$m and 160~$\mu$m and the Spectral and Photometric Imaging REceiver (SPIRE; \citealt{Herschel_SPIRE}) at 250~$\mu$m, 350~$\mu$m and 500~$\mu$m. The Full Width at Half Maximum (FWHM) of the PSF is 5\farcsecp2, 7\farcsecp7, 12\farcsecp0, 18\farcsecp2, 24\farcsecp9, 36\farcsecp3 in these bands, respectively. The details of all PACS and SPIRE data reduction steps, including error estimation, can be found in \citet{Remy2013}.

The PACS observations were performed as four pairs of orthogonal scans covering an area of 24\arcmin$\;\times\;$24\arcmin. We used an adapted version of the standard script of v7.0 of the Herschel Interactive Processing Environment (HIPE; \citealt{HIPE}). The basic processing includes flagging bad or saturated pixels, converting the signal into Jy/pixel and applying the flatfield correction. Additionally, we systematically masked column~0 of all of the constituent 16$\;\times\;$16 matrices in the PACS array to avoid electronic crosstalk and we performed second level deglitching. The resulting Level~1 products were converted into maps with the pixel size of 2, 2 and 4~arcsec for the three PACS bands and processed with Scanamorphos \citep{Roussel2012}, which is particularly suitable for extended sources with low frequency noise. The integrated PACS fluxes of NGC~4449 are well represented by a 40~K blackbody and suitable colour corrections were applied accordingly \citep{Muller2011b}, where no additional correction factors are required to correct for the extended nature of the source \citep{Sauvage2011}. The quoted uncertainties include the calibration uncertainties at 5 per~cent in all bands (\citealt{Remy2013}, \citealt{Muller2011a}).

The SPIRE observations were performed as two orthogonal scans covering an area of 24\arcmin$\;\times\;$24\arcmin. The corresponding maps were reduced using a modified version of the SPIRE pipeline in HIPE. The steps up to Level~1 were identical to those in the original pipeline provided by the SPIRE Instrument Control Center (ICC). Additionally, residual baseline subtraction was performed by subtracting the median of the time-lines for each bolometer over the entire observation \citep{Pohlen2010}. This was followed by an iterative process to completely remove residual signals that appear as stripes in the maps \citep{Bendo2010}. The final map was constructed using Naive Mapper available in HIPE and calibrated for an extended source. A modified blackbody fit of the form $S_\nu \propto \nu^\beta B_\nu (T)$ yields $\beta\approx2$ globally for NGC~4449 \citep{Remy2013}, which was used for colour correction of the integrated fluxes (SPIRE Observers' Manual; \citealt{Valtchanov2011}). The uncertainties include the revised overall calibration uncertainties at 7 per~cent in all SPIRE bands \citep{Griffin2011}.

\subsection{2MASS, \WISE\ and \Planck\ catalogue data}

Global 2MASS photometric measurements for NGC~4449 were taken
from columns {\tt j\_m\_ext}, {\tt h\_m\_ext} and {\tt k\_m\_ext}
of the 2MASS All-Sky Extended Source Catalog, and were
subsequently converted to flux densities using the zero points tabulated by \citet{Cohen2003}.
\WISE\ photometry was taken from columns {\tt gmag} and {\tt gerr}
of the \WISE\ All-Sky Source Catalog. Colour correction was applied
using interpolated correction factors suitable for a source emitting
as $\nu^{-1.18}$ across the four \WISE\ bands.
\Planck\ photometry was taken from column {\tt GAUFLUX(\_ERR)}
of the Early Release Compact Source Catalogue. Similarly, the interpolated
colour correction factors used were suitable for a source emitting
as $\nu^{2.80}$ across the \Planck\ bands centred at 350,
550 and 850~$\mu$m.

\section{Modelling method}\label{sec:method}

\subsection{Overview}\label{sec:method_overview}

The numerical code MOCASSIN (MOnte CArlo SimulationS of Ionized Nebulae;
\citealt{Ercolano2003,Ercolano2005,Ercolano2008}) was originally intended
as a tool to construct realistic models of photoionized nebulae.
It allows arbitrary three-dimensional geometries, separate for gas
and dust, multiple ionizing sources emitting with a given input spectrum,
variable gas chemistry, multiple dust species and arbitrary dust grain size
distributions. Given these input parameters, the code self-consistently
solves the radiative transfer in the coexisting gas and dust phases
and calculates the ionization degree, electron temperature and dust
grain temperature at every grid cell, and the overall emergent SED of the gas and dust.

In the sections that follow we will describe how MOCASSIN (v2.02.70) can
be used to model large physical systems, such as galaxies.
Our model is constructed `bottom-up', where as many parameters
as possible are fixed a priori based on observations.
The simulations are set up with the empirical elemental
abundances and the observed radial gas distribution fixed.
The input stellar spectrum corresponding to a mixture of
stellar populations, scaled by the stellar luminosity \Lstar, and the dust to
gas mass ratio (\DGR) are free parameters.
Based on these input parameters, which form a theoretical model of a galaxy,
MOCASSIN produces a predicted low-resolution SED and a full set of predicted
emission line intensities. If these predictions agree with observations,
the original parameter set can be regarded as a true representation of
the galaxy under the assumptions made. Otherwise, individual
parameters can be adjusted and simulations repeated iteratively,
as summarised schematically in Fig.~\ref{fig:n4449_mocassin_method}.

\input{./figures/figure.mocassin.method.tex}

In constructing our MOCASSIN model of NGC~4449 we have made simplifying assumptions about
the geometry, gas density distribution, elemental abundances,
dust composition and dust grain size distribution.

We have also made assumptions about the star formation history. In Section~\ref{sec:assumptions_sfh}
we discuss a general picture of the stellar populations in NGC~4449,
obtained using the spectral decomposition code STARLIGHT. In Section~\ref{sec:lstar}
we characterise the youngest stellar population based
on observational constraints. Finally, in Section~\ref{sec:criteria_episodes}
we fit the older populations taking into account the
constraints from earlier sections.

The fitting of the star formation history and dust is performed almost
simultaneously for self-consistency. Therefore, the young stellar population, the episodes
of star formation and dust are never modelled on their own. The youngest stellar
population is fitted together with a representative sample of older populations, and
the final dust content is determined relatively early in the process to ensure self-consistency.

The assumptions listed below form basis of the iterative scheme described in detail in Section~\ref{sec:method_all_variables_and_convergence}.

\subsection{Gas density distribution}\label{sec:assumptions_gas_density}
Spherical symmetry is assumed. Although MOCASSIN is a fully
three-dimensional code, this capability is not exploited at this time
due to technical limitations (see Section~\ref{sec:numerical_setup}).
However, the method presented in this paper can be very easily adapted for fully
three-dimensional modelling and the first step towards such modelling is discussed
in Section~\ref{sec:results_distributed}.

We used the H~{\sc i} radial profile of
\citet{Swaters2002} to approximate $n_\textrm{H}(r)/\textrm{cm}^{-3}$ as
four second-order polynomials in the form $ar^2+br+c$ for radii up to $3.3\;\textrm{kpc}$ (3\farcmin0),
corresponding to the extent of the galaxy in the FIR (Fig.~\ref{fig:n4449_5imgs}).
Although NGC~4449 is highly irregular, for simplicity we treat the
star formation regions cumulatively, and based on the optical
image (Fig.~\ref{fig:n4449_5imgs}) we assume that all gas is ionized within a
distance of $\sim\,$0.4~kpc from a single ionizing source.
The total observed H~{\sc i} mass for NGC~4449 is $2$--$2.5\times10^9\;\textrm{M}_\odot$ 
\citep{Swaters2002,Bajaja1994}, which is distributed between a system of
streamers of mass $9\times10^8\;\textrm{M}_\odot$ and the central region with diameter
24--35~kpc and mass $1.1$--$1.25\times10^9\;\textrm{M}_\odot$ \citep{Hunter1998,Hunter1999}.
The H~{\sc i} mass within the radius of
$3.3\;\textrm{kpc}$ derived using our formulation is $5.5\times10^8\;\textrm{M}_\odot$,
which is consistent with the estimated total of $1.1$--$1.25\times10^9\;\textrm{M}_\odot$ for the inner system
given the radial surface density profile of \citet{Swaters2002}.

The mass of H$_2$, based on $L_{\textrm{CO(1--0)}}=8.4\times10^6\;\textrm{K}\,\textrm{km}\,\textrm{s}^{-1}\,\textrm{pc}^2$ 
(corrected for $D = 3.8\;\textrm{Mpc}$; \citealt{Bottner2003}) and assuming
$\alpha_{\textrm{CO}} \sim 10$--$20\;\textrm{M}_\odot\,(\textrm{K}\,\textrm{km}\,\textrm{s}^{-1}\,\textrm{pc}^2)^{-1}$
(cf. \citealt{Sandstrom2012}, \citealt{Schruba2012} and references therein), can be estimated
at $\sim\;$$0.8$--$1.7\times10^8\;\textrm{M}_\odot$. However, we note that
CO-to-H$_2$ conversion factors for low metallicity objects are very uncertain,
and therefore we include no estimates of H$_2$ mass in our gas mass totals.

We assumed that all gas exists
in small clumps described by the filling factor $\epsilon$,
which is the same at all radii.
Considering a Str\"{o}mgren sphere of radius 0.4~kpc with $L_{\textrm{H}\alpha}$ given
by \citet{Hunter1999} and assuming that the average
electron density is equal to the central $n_\textrm{e}$ given by \citet{Martin1997},
we derive an initial estimate of $\epsilon \sim 0.003$.
This is revisited in Section~\ref{sec:lstar}.
Using similar assumptions and the formalism of \citet{Barlow1987} we
estimate that the mass of ionized gas is $\sim\;$$1.5\times10^6\;\textrm{M}_\odot$.

\subsection{Elemental abundances}
Elemental abundances are assumed to be constant throughout the galaxy.
The abundances were taken from \citet{Vigroux1987}, except for carbon,
neon and sulphur, where the Large Magellanic Cloud (LMC) ratios for C/O, Ne/O and S/O were assumed \citep{Russell1992}.
Although the sophisticated framework offered by MOCASSIN can also be
used to determine elemental abundances \citep{Ercolano2010},
in this paper MOCASSIN is not used in this context.

\subsection{Star formation history} \label{sec:assumptions_sfh}

Star formation histories are difficult to constrain
because of observational limitations
(e.g. instrumental sensitivity), but also
because of the more fundamental age-metallicity degeneracy
(e.g. \citealt{Worthey1994,Ferreras1999,Ferreras2003}).
However, a galactic spectrum
contains age-sensitive features (e.g. the H$\beta$ absorption line;
\citealt{Worthey1997}) and metallicity-sensitive
features (e.g. the [MgFe] index; \citealt{Gonzalez1993}, \citealt{Thomas2003}), which
can be useful in breaking this degeneracy.
At lower spectral resolutions the absorption features
produced by massive stars, for example Balmer lines,
can be weakened by coincident emission within
a starburst galaxy. However,
higher-order Balmer lines are less prone to this effect
because the strengths of corresponding nebular Balmer emission lines
strongly decrease with decreasing wavelength \citep{GonzalezDelgado1998}.
Therefore, a combined analysis of higher-order Balmer lines and other
tracers may give an indication of the ages and metallicities of the
dominant underlying stellar populations in a galaxy.
To obtain a range of possible ages and metallicities
in NGC~4449 we used the spectral fitting code STARLIGHT
(v04; \citealt{STARLIGHT1}, \citealt{STARLIGHT2})
and the global optical spectrum acquired by \citet{Kennicutt1992}.
This analysis will provide a first indication of the
star formation history, which will form basis for
more detailed stellar population fitting in Section~\ref{sec:criteria_episodes}.

From the observed optical spectrum, sampled every 2~\r{A}, we selected the range 3700--4900~\r{A}
for fitting, with the contaminating Balmer emission lines H$\beta$--H$\eta$, as well as the
unresolved [O~{\sc ii}]$_{\lambda\lambda3726,9}$ doublet and [Ne~{\sc iii}]$_{\lambda 3869}$,
masked out. We adopted 280~$\rm{km}\,\rm{s}^{-1}$ for the global recession velocity by
direct measurement from the observed spectrum, but the recession velocities are generally
lower and vary as a function of position \citep{Valdez-Gutierrez2002}.
The observed internal velocity dispersion, on the other hand,
is $\sim\,$30~$\rm{km}\,\rm{s}^{-1}$
\citep{Fuentes-Masip2000}. Since the grating selected by
\citet{Kennicutt1992} offered a resolving power of only
R$\;\sim\;$900, the velocity dispersion cannot be resolved by the observations
and the code cannot take advantage of the more suitable high-resolution
stellar spectra for fitting.
Nevertheless, the distribution of stellar populations
resulting from spectral fitting performed here
is not sensitive to the assumed internal velocity dispersion.

We used the synthetic single stellar population (SSP)
spectra of \citet{BC2003} with R$\;\sim\;$2000, which were
generated with the stellar initial mass function (IMF) of \citet{Chabrier2003}
and provided as part of the STARLIGHT package. This assumed IMF is very similar to
and consistent with the IMF of \citet{Kroupa1991,Kroupa2001} adopted in this work.
STARLIGHT was run with 12 sets of 5--99 stellar templates, or `bases',
to probe the star formation history at different sampling resolutions of age and metallicity.
The lower and upper limits of the stellar ages were 1~Myr and 12~Gyr and
1--3 metallicities, corresponding to 0.2--1.0 $\textrm{Z}_\odot$, were tested at a time.
One of the fits is presented in Fig.~\ref{fig:n4449_STARLIGHT_example}.

Our fits showed a significant spread in age--metallicity pairs at high
resolutions of age and metallicity. Therefore, in Table~\ref{tab:starlight} we present
a summary of the fitted populations, grouped into `young', `intermediate' and `old'
and based on results from all fits.
Table~\ref{tab:starlight} shows that the most metal-poor populations
are relatively well constrained and
represent almost all of the stellar mass in NGC~4449. We find no evidence 
for a significant mass of stars with metallicities above $0.2\textrm{Z}_\odot$
amongst the old ($\gtrsim\;$1~Gyr old) population.
Based on our results, we estimate the first onset of star formation
at approximately 12 Gyr ago. Because STARLIGHT tended to select the
oldest available template for the oldest population, we emphasise that this
result is approximate and onset epochs ranging from 10 Gyr ago
to the age of the Universe should be considered equally plausible.
Our results also consistently point to an age of 400$\,\pm\,$100 Myr
as being representative for the intermediate population.

\input{./figures/figure.STARLIGHT.example.tex}
\input{./tables/table.STARLIGHT.results.tex}

These results are supported by the resolved
stellar population study of \citet{Annibali2008}, suggesting continuous
star formation activity for at least 1~Gyr,
and by the results of \citet{Bothun1986}, who
estimated that the underlying old stellar component has a mean age of 3--5~Gyr.
A very young population of stars, on the other hand, is
inferred from the presence of Wolf-Rayet stars \citep{MartinKennicutt1997}.

Therefore we assume that star formation started not later than $\sim\,$4~Gyr ago
and continued until very recently or continues to the present day.
We also assume three continuous star formation episodes.
This assumption reduces the complexity of the model,
and, at the same time, hints at a possible star formation history.
The continuous-episode approach can be viewed as a generalisation of
a starburst approach, which is discussed in Section~\ref{sec:results_preliminary}.

\subsection{Interstellar dust}\label{sec:assumptions_dust_composition}

\subsubsection{Distribution and composition}\label{sec:assumption_evolution_model}

The dust distribution is assumed to follow that of the gas,
with the dust to gas mass ratio (abbreviated here as DGR or \DGR)
initially kept constant throughout the galaxy
(however see, e.g. \citealt{Bianchi2008}, \citealt{Baes2010}, \citealt{MacLachlan2011},
\citealt{SchechtmanRook2012}, \citealt{deLooze2012}).
We assumed two dominant dust species:
amorphous carbon \citep{Hanner1988} and silicates \citep{Laor1993}.
To limit the number of free variables in the model, we further assume
a constant mass ratio of carbonaceous dust to silicaceous dust
(hereafter referred to as the carbon-to-silicate ratio) as a function of position.
In what follows, we intend
to illustrate explicitly the evolution of the carbon-to-silicate 
ratio for a galaxy continuously forming stars
to allow us to fix this ratio for NGC~4449.
We refer the reader to \citet{Dwek1998}, \citet{Morgan2003}, \citet{Galliano2008a}
and \citet{Dwek2011a}
for more complete studies of dust evolution in the ISM.

The refractory dust in the ISM is believed to form in the
outflows from asymptotic giant branch stars (AGB stars; e.g. 
\citealt{Matsuura2009} for the LMC), in
type~II core-collapse supernova (CCSN) ejecta (e.g. a multi-epoch study of SN~2004et by \citealt{Fabbri2011}),
and has also been suggested to form in molecular clouds \citep{Draine2009}.
\citet{Dwek1998} has suggested that low-mass AGB stars are the main source
of carbon dust, while type~II SNe are the
main source of silicate dust. Thus, the
relative abundance of carbon and silicate dust depends on the dynamics
of dust formation by AGB stars and supernovae, and on dust destruction
processes and thus on the lifetimes of individual dust species.

Our calculations assume the dust yields of
\citet{Ventura2012} for AGB stars,
a constant and continuous star formation
activity (Section~\ref{sec:assumptions_sfh}, above), a constant metallicity, 
the initial mass function (IMF) of \citet{Kroupa1991,Kroupa2001}
for stellar masses $1.5\;\textrm{M}_\odot<M<40\;\textrm{M}_\odot$,
and ignore dust destruction processes including consumption by ongoing star formation.

It is informative to first consider a flat distribution of
yields across all SN masses and the carbon-to-silicate ratio
inferred for SN~1987A by \citet[ Model 1]{Matsuura2011}.
Fig.~\ref{fig:n4449_dust_evolution_model}
shows the carbon-to-silicate ratio for a galaxy actively forming
stars for a continuous period of 4~Gyr. Since the data of \citet{Ventura2012}
do not include AGB stars of mass lower than 1.5 $\textrm{M}_\odot$,
our models cannot be reliably extrapolated beyond the first 4~Gyr.
At early times, the carbon-to-silicate ratio is
fixed by the assumed dust composition of SN ejecta. When the SN dust 
dominates the global dust budget, the ratio is not expected
to change significantly with time (Fig.~\ref{fig:n4449_dust_evolution_model}, left, 
red solid line). If AGB dust dominates dust production, the
carbon-to-silicate ratio initially decreases after $\sim\,$40~Myr due to
the predicted
injection of silicates by super-AGB stars. After $\sim\,$400~Myr, less massive
AGB stars begin injecting carbonaceous dust leading to carbon-dominated
ISM dust after $\sim\,$1~Gyr (Fig.~\ref{fig:n4449_dust_evolution_model}, left,
blue dotted line).

A more realistic model includes variations in
dust mass production as a function of CCSN progenitor mass, based
on CCSN element yields of \cite{Woosley1995}.
Since the metallicity assumed by \citet{Ventura2012} is $Z/\textrm{Z}_\odot=0.05$,
we chose the closest $Z/\textrm{Z}_\odot=0.1$ $^{56}$Ni-producing models
from \cite{Woosley1995} for all stellar progenitor masses,
thus ignoring fallback of heavy elements \citep{Moriya2010}.
We further scaled the CCSN elemental yields by a
constant factor of 0.2 to bring the predicted dust masses into approximate
agreement with the dust masses inferred from a range of supernovae
(\citealt{Wesson2010}; \citealt{Temim2006}; 
\citealt{Barlow2010}; \citealt{Matsuura2011}; \citealt{Gomez2012}). This scaling factor
may be physically interpreted as the adopted condensation efficiency
of dust. As amorphous olivine is observed to
dominate the silicate mass in the diffuse ISM \citep{Kemper2004},
we assumed silicate stoichiometry of MgFeSiO$_4$ for
calculations of silicate dust masses.

The results in Fig.~\ref{fig:n4449_dust_evolution_model} (right) show
that the carbon-to-silicate
ratio increases with time, owing to more carbon-rich dust being
produced by lower mass SNe. This
is strengthened at later times by carbon-rich dust contribution from the
less massive AGB stars.

\input{./figures/figure.n4449.dust.evolution.tex}

In this model
a single metallicity of $Z/\textrm{Z}_\odot$$\,\sim\,$$0.1$ was assumed for the dust.
However, the yields of \citet{Woosley1995} suggest that as
metallicity increases, the relative production of carbon
by CCSNe also increases. The opposite trend is suggested 
for AGB stars. In low metallicity environments
AGB stars are expected to produce less silicates,
while maintaining the same or an
enhanced production of carbon dust \citep{Sloan2006,Sloan2012} than
in higher metallicity environments.
Since these trends operate on different time-scales, their combined
effect is likely to result in a carbon-to-silicate profile which is
more steeply increasing than the one shown in Fig.~\ref{fig:n4449_dust_evolution_model}.

Therefore, given our assumptions about metallicity, the limited period of
modelled star formation activity, and
the uncertainty of the degree of dust enrichment by SNe,
our final carbon-to-silicate mass ratio of 0.23 (Fig.~\ref{fig:n4449_dust_evolution_model}, right) is likely
to represent the lower limit of relative carbon content in a
continuous star formation scenario. NGC~4449, with
$\log(\textrm{O}/\textrm{H})+12=8.23$, is similar in terms of metallicity
to the LMC \citep{Russell1992},
whose dust content suggests a slightly higher carbon-to-silicate mass ratio of 0.33
\citep{Weingartner2001}.
If our model is allowed to evolve beyond the first 4~Gyr, the relative carbon content
in NGC~4449 would be expected to match or exceed the LMC value, depending on the
age of NGC~4449. This observation may warrant the use of metallicity-based assumptions
about the carbon-to-silicate ratio in future studies.
Therefore, given that our spectral fitting results discussed in Section~\ref{sec:assumptions_sfh}
point to an early onset of star formation, we adopt the LMC 
carbon-to-silicate mass ratio of 0.33.

\subsubsection{Grain size distribution}
While the absence of distinct silicate emission features
in the \Spitzer\ Infrared Spectrograph (IRS) spectra \citepalias{Karczewski2013_OBSERVATIONS}
may indicate a small abundance of Si, our chemical evolution models
(Section~\ref{sec:assumption_evolution_model}, above)
predict a non-negligible mass of silicate dust.
If all dust is assumed to reside in a thin layer
around the central ionizing source, this would impose an
upper limit of 20 per~cent on the mass fraction of Si in `ultrasmall' silicate grains ($<\,$15~\r{A}; \citealt{LiDraine2001}) and
suggest that the contribution of hot silicate dust to the global dust budget
is relatively small (e.g. \citealt{Smith2010}).
However, NGC~4449 is a face-on irregular galaxy and it is reasonable to assume,
to a first-order approximation, that the radial distribution of dust follows
the radial distribution of gas.
Therefore, the absence of distinct silicate emission features in the global IRS spectrum
may indicate effects resulting from geometry, for example
extinguishing the emission from small grains by self-shielding.

We assume that the dust grain size distribution
follows $n(a) \propto a^{-3.5}$ \citep{MRN77} and
we adopt the same range of grain sizes $a$ between 0.005 $\mu$m and 0.25 $\mu$m for both
amorphous carbon and silicates.

\subsubsection{Polycyclic aromatic hydrocarbons}\label{sec:assumptions_pahs}

Prominent emission features in the mid-infrared, also
known as the unidentified infrared bands (UIBs), are visible
in the spectra of many types of objects, and are also detected
in the \Spitzer\ spectrum of
NGC~4449 \citepalias{Karczewski2013_OBSERVATIONS}.
They are believed to originate from the C--H and C--C vibrations
of large organic molecules, or polycyclic aromatic hydrocarbons
(PAHs; e.g. \citealt{Allamandola1985}, \citealt{Peeters2002}, \citealt{Bauschlicher2009}).
The relative strengths of these emission features are believed to reflect
the particle size distribution and the local physical environment
of the PAH molecules (e.g. \citealt{Galliano2008b}, \citealt{Bauschlicher2009}).

We assumed the mathematical description of pure ionized PAHs
from \citet{Draine2007} with the grain size
distribution of \citet{Weingartner2001} for grain sizes
of 3.5--30 \AA. The PAH masses
in this approach are not constrained
by abundances (cf. \citealt{Zubko2004}).
Since MOCASSIN currently allows only one
grain size distribution per simulation,
the PAHs have to be modelled separately.

\subsection{Numerical setup}\label{sec:numerical_setup}

A series of preliminary simulations showed that the best compromise between
high spatial resolution and reasonable computing time is obtained for a grid
of $80\times80\times80$ cells, corresponding
to 0.5 million grid cells. In a spherically symmetric setup, MOCASSIN
uses a Cartesian $x,y,z$ grid, with
cells populating one octant of a sphere resulting in an eight-fold reduction of computing
time for a given spatial resolution. The grid cells were not equal in size, and
the resolutions used were 15~pc, 50~pc and 185~pc for the inner 0.5~kpc region,
intermediate radii up to 2.5~kpc, and for outer radii, respectively. 
The number of grid cells and the
resolutions ensure adequate sampling of
the gas density profile (Section~\ref{sec:assumptions_gas_density}) as 
a function of radius. A single ionizing source, describing the stellar
content of the entire galaxy, was placed at the origin of the coordinate system.
The physical inner and outer radii of the galaxy were defined as 1~pc and 3300~pc.

The computation time for a single gas and dust simulation
resulting from the setup described in this paper
was $\sim\,$9~h using 8~CPUs with 8~GB of RAM per CPU. 

\subsection{Variables and convergence criteria}\label{sec:method_all_variables_and_convergence}

As described in Section~\ref{sec:method_overview} and
shown in Fig.~\ref{fig:n4449_mocassin_method},
the three variables,
the global bolometric luminosity \Lstar,
the DGR and the assumed star formation history (SFH),
are iteratively adjusted until all
observables are well matched. Of these, \Lstar\ is the most
independent and can be fixed
at early stages of the modelling.

The DGR and the SFH
are not independent, as a higher dust content
increases the extinction at the UV and optical wavelengths and
therefore requires more ionizing photons to maintain
the observed continuum levels. The SFH
in itself is a multi-dimensional function, which
can be explored to reach the most
probable solution.
We use the general picture of the SFH emerging from
Section~\ref{sec:assumptions_sfh}
as a constraint for constructing trial three-episode
star formation scenarios.
The steps described below may be viewed
as an attempt to minimise
a multi-dimensional mathematical
expression by examining its partial derivatives locally.

\subsubsection{The bolometric stellar luminosity \Lstar}\label{sec:lstar}

A large fraction of all the energy 
emitted by a starburst galaxy
comes from the youngest and most massive stars.
These stars contribute to the 
observed UV and optical continuum and
govern the observed global ionization
states of species present in the ISM. The same radiation
field is also partially absorbed
by the interstellar dust and re-radiated
thermally in the IR. The degree of ionization
of nebular species and the degree of
reprocessing of radiation
are also dependent on the clumpiness of the ISM.

Therefore, by focusing initially
on the youngest stellar component and considering a representative
set of star formation scenarios it is possible not only {\it (i)} to 
further simplify the SFH by constraining this youngest component,
but also {\it (ii)} to fix the filling factor $\epsilon$, {\it (iii)} to fix \Lstar\ and {\it (iv)}
to provide an initial estimate for the DGR.
Two main constraints, namely, the absolute level of
the UV continuum and the `recent' 
star formation rate (SFR) as traced by H$\alpha$ or a similar tracer,
must both be matched by the total stellar luminosity and the
average age of the youngest population given the adopted filling factor.

The `recent' star formation rate for NGC~4449
is estimated at $\sim\,$0.5~$\textrm{M}_\odot\,\rm{yr}^{-1}$
\citep{Hunter1986,Hunter1999} for a Salpeter IMF (\citealt{Salpeter1955}; see discussion
in Section~\ref{sec:imf}). Since
\citet{Dopita2006} showed that for a SSP all ionizing
photons are emitted within 10~Myr,
we adopt 10~Myr, the lifetime
of O- and B-type stars with masses $\gtrsim\,$13 $\textrm{M}_\odot$,
as a representative period
of continuous star formation for the youngest stars.

The trial star formation histories
(here also referred to as `scenarios') were
generated using STARBURST99
(\citealt{Leitherer1999}; v6.0.2) with 
the IMF of Kroupa \citep{Kroupa1991,Kroupa2001},
along with the
Pauldrach/Hillier model atmospheres \citep{SmithNorrisCrowther2002},
the updated Padova AGB tracks \citep{Vazquez2005} and
an assumed single constant metallicity of $Z/\textrm{Z}_\odot=0.4$ (cf. Table~\ref{tab:starlight}).

For a representative set of scenarios consisting of three star formation episodes,
we first varied the degree of clumpiness, represented by $\epsilon$,
to allow enough gas to be ionized and produce the observed nebular emission
line intensities \citep{Kobulnicky1999} to within a factor of a few.
We found that $\epsilon = 0.033$ is most representative, which corresponds
to an average $n_{\textrm{e}}$ of $41\;\textrm{cm}^{-3}$ within the modelled ionized
region. This choice is discussed further in Section~\ref{sec:results_ionisation}.

We constructed a grid in parameter space formed by
a representative set of scenarios consisting of
three star formation episodes, a range of SFRs of the
youngest population which is continuously forming
stars for a period of 10~Myr, and a small range
of initial values for \Lstar\ and the DGR.
The predictions were then
compared with observations. In particular, 
\begin{enumerate}
\item the predicted UV continuum level
was compared with the observed SED (Table~\ref{tab:photometry_table}),
\item the estimated total number of ionizing photons, $Q(\textrm{H}^0)$,
was compared with that derived from the integrated H$\alpha$ luminosity \citep{Hunter1999} 
and
\item the predicted nebular emission line
fluxes were compared with the measurements of \citet{Kobulnicky1999}.
\end{enumerate}

The best-fitting models were found by minimising residuals between the model
and the observations in a way similar to $\chi^2$ minimisation.
At this stage of modelling more weight was
given to line intensities and to the level and shape of the continuum in the UV.
To reduce the extent of the parameter space and the overall computation time,
the iterative scheme in Fig.~\ref{fig:n4449_mocassin_method}
was not allowed to advance automatically.
Instead, after each iteration, candidate parameter sets
were inspected to identify trends in each variable.
The most promising parameter combinations were then manually expanded into
a higher-resolution parameter space and iterated according to Fig.~\ref{fig:n4449_mocassin_method}.

Our results suggest an on-going
star formation activity with recent SFR$\;\approx\;$0.28~$\textrm{M}_\odot\,\rm{yr}^{-1}$,
\Lstar$\;\approx\;$$5.7\times10^9\,\textrm{L}_\odot$ and an initial
DGR$\;\sim\;$1/250, thus allowing us to fix the youngest population
and \Lstar. The empirical \Lstar, computed by
integrating measurements in Table~\ref{tab:photometry_table}
between 0.15~$\mu$m and 3.5~$\mu$m, is $3.1\times10^9\,\textrm{L}_\odot$.
This suggests that a significant fraction of the total luminosity
originates from the recently formed massive stars.

\subsubsection{The initial mass function}\label{sec:imf}

The recent SFR of 0.28 $\textrm{M}_\odot\,\rm{yr}^{-1}$,
suggested by our initial models,
is in good agreement with previous estimates (see below), when
the revised distance of 3.8~Mpc \citep{Annibali2008}
and differences in the assumed IMFs are taken into account.

The Salpeter IMF, often
used in the formulations of the
SFR (e.g. \citealt{Kennicutt1998}), is defined as
$N(M)\;\mathrm{d}M \propto M^{-2.35}\;\mathrm{d} M$ \citep{Salpeter1955}.
However, the Kroupa IMF, which is assumed in this work,
is defined in two intervals,
with a flatter distribution of the number of stars
at lower stellar masses. Therefore,
the Salpeter IMF will overestimate the number of lower-mass stars compared
to the Kroupa IMF. This will in turn overestimate the total
mass of stars resulting in a higher prediction of the SFR.

We generated distributions of stellar masses normalised to give
the same number of stars at a stellar mass of 25 $\textrm{M}_\odot$,
which we assumed to be representative of the population
generating the ionizing photons and giving rise to the observed
H$\alpha$ emission. We computed the total mass of stars from
$N(M)M\;\mathrm{d}M$ for $0.1 \leq M/\textrm{M}_\odot \leq 100$, which
yielded a factor of 1.54 difference between the two IMFs.
Assuming that the integrated luminosity of a galaxy scales
linearly with the total mass, the SFRs based on the Salpeter IMF
are therefore likely to be a factor of $\approx\,$1.5 higher than if the Kroupa IMF were used.
This result is also evident from the results of \citet{Dwek2011b}, who
tabulate the masses of all stars born per one supernova event for
a range of stellar IMFs.

Overall, our recent SFR of 0.28~$\textrm{M}_\odot\,\rm{yr}^{-1}$
is in good agreement with the distance-corrected and IMF-corrected
estimates of 0.22--0.33 $\textrm{M}_\odot\,\rm{yr}^{-1}$ \citep{Hunter1986},
0.30 $\textrm{M}_\odot\,\rm{yr}^{-1}$ \citep{Hunter1999},
and 0.21 $\textrm{M}_\odot\,\rm{yr}^{-1}$ (based on high ionization potential neon lines; \citetalias{Karczewski2013_OBSERVATIONS}).

\subsubsection{Three episodes of star formation}\label{sec:criteria_episodes}

To constrain \Lstar\ and the youngest stellar population,
we assumed a representative older stellar component.
Since the contribution of stars older than
$\sim\,$10~Myr to the ionization structure,
to the number of ionizing photons, or to the far-UV
continuum, is very small, the older populations
can be studied in more detail without affecting the validity of
our findings from Section~\ref{sec:lstar}.

The shape of the observed SED from the
UV to the near-infrared (NIR) can be fitted by varying the
lengths and characteristic epochs determining the remaining episodes
of continuous star formation. Initially, we tested a set
of 36 scenarios, which consisted of all combinations of:
\begin{enumerate}
\item three epochs corresponding to the onset of star formation (4, 8 and 12~Gyr ago),
\item four epochs corresponding to the transition between the first (old)
episode and the second (intermediate) episode of star formation (100, 200, 400 and 1000~Myr ago),
\item three star formation rates ranging from 0.05 $\textrm{M}_\odot\,\rm{yr}^{-1}$
to 0.15 $\textrm{M}_\odot\,\rm{yr}^{-1}$ further defining the second episode.
\end{enumerate}
The star formation activity during the third (young) episode
remained fixed in both duration and magnitude, and \Lstar\ represented the
luminosity of all stellar components, as described in Section~\ref{sec:lstar}.
Therefore, the star formation rate of the first (oldest) episode was
not an independent variable, but depended on \Lstar\ and the durations
and star formation rates in items 1--3, for each scenario.

The resulting low-resolution SEDs were compared with the observed
SED to select the best matches. Afterwards, following the scheme
shown in Fig.~\ref{fig:n4449_mocassin_method}, the transition epochs
and the star formation rates inferred from the best matches were
expanded into a narrower, but higher-resolution parameter space,
analogous to conditions in items 2--3.

The criteria for finding the best-fitting combinations
of stellar populations were similar to those used to find \Lstar,
where we minimised residuals between the model
and the observations. At this stage, more weight was given to the
number of ionizing photons $Q(\textrm{H}^0)$, the line intensities
and the level and shape of the entire stellar continuum from
the UV to the NIR.
As before, to reduce the extent of the parameter space and the overall computation time,
the iterative scheme in Fig.~\ref{fig:n4449_mocassin_method}
was not allowed to advance automatically.
Instead, after each iteration, candidate parameter sets
were inspected {\it (i)} to identify trends in each variable and
{\it (ii)} to verify if the fit could be improved while
still satisfying the observed number of ionizing photons $Q(\textrm{H}^0)$.
The most promising parameter combinations were then manually expanded into
a higher-resolution parameter space.

\subsubsection{A two-zone solution for \DGR\ and modelling PAHs}\label{sec:variables_mdmg}\label{sec:results_PAH}

NGC~4449 is modelled as a centrally-concentrated, spherically-symmetric
galaxy, even though its morphology is highly irregular (cf. Fig.~\ref{fig:n4449_5imgs}).
Therefore, spherical
symmetry inevitably misrepresents the interstellar
radiation field (ISRF) within the galaxy. The
single central source is much more luminous
than any individual source in the real galaxy, as it must deliver enough
energy to the correct mass of gas and dust to reproduce the observed
strengths of the emission lines and the observed
SED. Consequently, the conditions in the immediate proximity of the
single central source result in unrealistically high ionization
of the gaseous species and vaporisation of any existing dust grains.
This effect is discussed in more detail in Section~\ref{sec:results_ionisation}.

In order to fit the dust emission in the FIR,
we divided the galaxy into two zones: an inner zone with a lower
DGR and an outer zone with a higher DGR, which ensures
realistic absolute masses of hot dust near the centre.
For technical reasons this was preferred to an alternative approach, in which
the radial distribution of dust density is
defined directly through the MOCASSIN input parameter {\tt Ndust}.
We found that for larger systems, such as
galaxies, defining dust distribution through
{\tt MdMg} gives best performance.

The best-fitting models were repeated with pure ionized PAH dust
and the PAH grain size distribution (Section~\ref{sec:assumptions_pahs}), and compared
with the observed global SED (Table~\ref{tab:photometry_table})
and the global \Spitzer/IRS spectrum \citepalias{Karczewski2013_OBSERVATIONS}
to estimate the total mass of PAH molecules.
In the final step, the resulting spectra were combined with the corresponding
amorphous carbon and silicate models.

\section{Results and Discussion}\label{sec:results}

\subsection{Preliminary models}\label{sec:results_preliminary}

Our preliminary models included both starburst-only scenarios and continuous
star formation scenarios. In the starburst-only scenario, a fit satisfying
all observational constraints
can be obtained for three representative starbursts occurring
3~Myr ago, 100~Myr ago and 4~Gyr ago, contributing to the total stellar mass in a ratio
1:200:1000. These models were later generalised by replacing starbursts
with longer star formation episodes. For a scenario
assuming two continuous star formation episodes the
inferred star formation rates were
$\sim\,$0.25 $\textrm{M}_\odot\,\rm{yr}^{-1}$ and $\sim\,$0.09 $\textrm{M}_\odot\,\rm{yr}^{-1}$
between $\sim\,$6~Gyr ago and $\sim\,$120~Myr ago, and $\sim\,$120~Myr ago and the present day,
respectively.
The star formation history in these models was constrained only by the observed SED and by
the observed emission line intensities. We note that
the representative populations and star formation rates are not too
dissimilar to the assumptions described in Section~\ref{sec:assumptions_sfh} and
the recent star formation rate of 0.28~$\textrm{M}_\odot\,\rm{yr}^{-1}$ (Section~\ref{sec:imf}).
Therefore, it may be argued that combining the
three convergence criteria, namely, {\it (i)} matching the observed SED,
{\it (ii)} matching the total observed rate of ionizing photons, $Q(\textrm{H}^0)$,
and {\it (iii)} matching the observed nebular emission line fluxes,
may give informative results, even for a relatively simplified model.

\subsection{Final model}

Fig.~\ref{fig:n4449_mocassin_bestfitting}
shows the best-fitting MOCASSIN model of NGC~4449
satisfying the observational constraints over the entire wavelength range
from the UV to sub-mm. The individual SEDs arising from the assumed three
continuous star formation episodes are also shown for comparison.
The input parameters and the results are summarised in Table~\ref{tab:MOCASSIN_inputs}
and the details of the best-fitting parameters
describing the star formation history are given in Table~\ref{tab:mocassin_SFH}.

\input{./tables/table.MOCASSIN.inputs.tex}
\input{./figures/figure.mocassin.bestfitting.tex}

The three episodes of star formation shown in Fig.~\ref{fig:n4449_mocassin_bestfitting}
suggest that most of the UV emission, and most of the
processed radiation emitted in the FIR, originates from the
intermediate-age stars of ages between 10 and $\sim$$\;400$~Myr. The
stars produced in the first episode, i.e., those older than $\sim$$\;400$~Myr,
dominate in the optical and in the NIR. The youngest stars of ages
less than 10~Myr contribute mainly in the far-UV.

Additionally, Fig.~\ref{fig:n4449_mocassin_variables} shows the
effects of varying the input parameters for the same best-fitting
model (red solid line). For example, the effect of varying \Lstar\ 
is a vertical shift of the predicted SED in the stellar part (Fig.~\ref{fig:n4449_mocassin_variables}, top panel)
and a corresponding change in $Q(\textrm{H}^0)$, as the IMF-weighted
distribution of stellar luminosities is scaled by a constant.

The predicted and observed global emission line intensities, relative to H$\beta$, are
shown in Table~\ref{tab:MOCASSIN_lines}.
The discrepancies between the predicted and the observed
nebular line intensities of some species, most notably
[S\two], result from the simplifying assumptions
about the geometry and the degree of clumpiness,
and are discussed in Section~\ref{sec:results_ionisation}.
Because the current version of MOCASSIN offers no treatment of
PDRs, the PDR line [O\one] in
Table~\ref{tab:MOCASSIN_lines} is underpredicted.
However, MOCASSIN is in the process of being expanded
to accommodate PDRs via the recently developed three-dimensional PDR code 3D-PDR \citep{Bisbas2012}.

\subsubsection{Stellar populations}

In our three-episode model a stellar mass of $\approx\,$$1\times10^9\;\textrm{M}_\odot$
is produced at rates 0.25--0.10, 0.14 and 0.28 $\textrm{M}_\odot\,\rm{yr}^{-1}$
over three periods, as shown in Table~\ref{tab:mocassin_SFH}.
The best-fitting SED in Fig.~\ref{fig:n4449_mocassin_bestfitting}
shows that this model does not 
fully account for the emission between 7000~\AA\ and 2~$\mu$m,
which may be a result of the assumption of only three episodes
of star formation
with a constant SFR for the duration of each episode. However,
we note that a non-trivial three-dimensional
dust distribution, where the degree of obscuration does not follow the
distribution of gas or stars, may have a significant effect on the observed attenuation
(e.g. \citealt{Witt1992}, \citealt{Baes2000}).

We found that the results are degenerate for the oldest
stars and therefore the first onset of star formation
4, 8 or 12~Gyr ago is equally plausible, resulting in a range of
predicted star formation rates. The spectral fits
performed with STARLIGHT and discussed in
Section~\ref{sec:assumptions_sfh} suggest an
onset 12~Gyr ago or earlier. For an onset
of star formation 12~Gyr ago, the
average star formation rate
for the duration of the first episode is
0.09 $\textrm{M}_\odot\,\rm{yr}^{-1}$.
Because the mass of
recently formed stars is small in comparison (cf. Tables~\ref{tab:starlight} and~\ref{tab:mocassin_SFH}),
it is possible that up to 95 per~cent of the total stellar mass $M_\star$ in NGC~4449
was produced at higher instantaneous rates
in an initial starburst.

The predicted onset of the second episode, 300--400 Myr ago,
coincides with a possible encounter with the galaxy DDO~125 400--600 Myr ago \citep{Theis2001}.
Given the simplifications of our model it would be
difficult to quantify the effects of this postulated encounter on star formation rates.
However, we note that our STARLIGHT fits (Section~\ref{sec:assumptions_sfh})
select a population $\sim\,$400~Myr old as being representative
for the intermediate-age populations. Similarly,
the three-episode MOCASSIN fits to the observed stellar SED converge
to 300--400 Myr ago as the transition epoch between the first
episode and the second.

Our model suggests that in
the last 10~Myr star formation has been taking
place continuously at an average rate of
0.28~$\textrm{M}_\odot\,\rm{yr}^{-1}$ (0.42~$\textrm{M}_\odot\,\rm{yr}^{-1}$ assuming
Salpeter IMF; Section~\ref{sec:imf}). This recent SFR,
obtained by fitting both the UV continuum and
the emission lines (Section~\ref{sec:lstar}),
is consistent with other estimates based on
the UV continuum emission (0.44~$\textrm{M}_\odot\,\rm{yr}^{-1}$; using
data in Table~\ref{tab:photometry_table} and following the formalism of \citealt{Kennicutt1998})
or emission line intensities (\citealt{Hunter1986}, \citealt{Hunter1999}, \citetalias{Karczewski2013_OBSERVATIONS}).

The FIR emission arises from dust heated by
stars and can also be used to estimate the recent SFR.
Integrating the best-fitting SED in Fig.~\ref{fig:n4449_mocassin_bestfitting}
between 8~and 1000~$\mu$m yields $I(\textrm{FIR}) \approx 3.6\times10^{-12}\;\textrm{W}\,\textrm{m}^{-2}$.
The corresponding recent star formation rate is 0.28
$\textrm{M}_\odot\,\rm{yr}^{-1}$ assuming a Salpeter IMF \citep{Kennicutt1998}, which
is significantly lower than the estimate of 0.44~$\textrm{M}_\odot\,\rm{yr}^{-1}$
based on the UV emission.
Indeed, the ratio SFR(FIR)/SFR(UV) for NGC~4449
based on the data in Table~\ref{tab:photometry_table}, and on
the fit presented in Fig.~\ref{fig:n4449_mocassin_bestfitting},
is 0.28/0.44$\;=\;$0.64, suggesting that approximately 35 per~cent
of the UV radiation does not contribute to the heating of dust.
This UV `leakage' (e.g. \citealt{Relano2012}) is observed
also in Haro~11 and NGC~4214 \citep{Cormier2012,Hermelo2013},
and may result from porosity of the ISM or 
different distributions for the stars and dust.

The low ratio SFR(FIR)/SFR(UV) may also suggest that,
globally, in NGC~4449 the dust is heated by younger stellar populations.
Similarly, a spatially-resolved study by \citet{Galametz2010}
suggests that the distribution of cooler dust within a different dwarf galaxy, NGC~6822,
may be correlated with star formation activity.
In spiral galaxies, on the other hand, several studies suggest that
the dust is heated by both the younger and the evolved stellar populations
(e.g. \citealt{Bendo2010}, \citealt{Boquien2011}, \citealt{Bendo2012a}).

\subsubsection{Interstellar dust}

We infer a dust mass of $2.9\pm0.5\times10^6\;\textrm{M}_\odot$ by making
assumptions about the distribution of dust and its characteristics:
the carbon-to-silicate ratio and the grain size distribution
(cf. Section~\ref{sec:assumptions_dust_composition} and Table~\ref{tab:MOCASSIN_inputs}).

As explained in Section~\ref{sec:variables_mdmg},
in modelling the dust emission within NGC~4449 we adopted two zones
with differing DGRs. Fig.~\ref{fig:n4449_mocassin_variables} (middle panel)
shows the two components, modelled simultaneously, in
green and blue. The combined effect of
the best-fitting DGRs and the adopted distribution of gas (Section~\ref{sec:assumptions_gas_density})
is a relatively constant distribution of
dust as a function of radius, varying by less than a factor of three between the centre and 
$r = 3.3\;\textrm{kpc}$,
and yielding $M_{\textrm{gas}}$/190 within the modelled region.
The need for two zones demonstrates that the actual distribution
of dust is significantly different from the distribution of gas, even for one-dimensional
azimuthally-averaged profiles.
Fig.~\ref{fig:n4449_mocassin_variables} also shows results for a smaller and a larger
overall DGR, suggesting that the uncertainty
in the best-fitting dust mass is $\sim\,$$0.5\times10^6\;\textrm{M}_\odot$.

The composition of dust has a major influence on the amount of attenuation
produced at UV and optical wavelengths, and therefore on the total
amount of dust required to reproduce the observed FIR emission.
Fig.~\ref{fig:n4449_mocassin_bestfitting} shows that the photometric
measurements agree with the model near the
prominent absorption feature at 2175~\AA, which
is attributed to carbon dust (e.g. \citealt{Fitzpatrick1986}).
The \Spitzer/IRS spectrum, on the other hand,
is dominated by the PAH emission features and shows no
discernible 10~$\mu$m silicate feature \citepalias{Karczewski2013_OBSERVATIONS},
and no additional underlying component
at 10~$\mu$m  was needed to fit the PAH emission.
Therefore, while appreciating that the assumed carbon-to-silicate ratio of
1:3 may be underestimated
(also see discussion in Section~\ref{sec:assumptions_dust_composition}),
we conclude that the adopted ratio is consistent with observations.
We find a contribution from PAHs to the total mass of dust of 2 per~cent, which is
consistent with the $M_{\textrm{PAH}}/M_{\textrm{dust}}$ ratios expected for
a galaxy with 1/3$\textrm{Z}_\odot$ \citep{Galliano2008a}.

For comparison, in Fig.~\ref{fig:n4449_mocassin_variables} (bottom panel, dotted line) we
show the SEDs corresponding to different compositions of dust for
the same total mass of dust. Additional carbon-rich dust (dotted line and dashed line)
produces more emission in the FIR. If such carbon-rich compositions were used in
fitting the SED, the overall required mass of dust would be lower.
More silicate-rich dust (dot-dashed line), on the other hand, would increase
the required mass of dust.
Dust mass estimates in the
literature range from $2.0\times10^6\;\textrm{M}_\odot$ \citep{Engelbracht2008}
to $3.8\times10^6\;\textrm{M}_\odot$ \citep{Bottner2003}, reflecting different properties
of dust adopted in interpreting the FIR emission.

The metallicity of NGC~4449 is similar to that of NGC~1705 and NGC~6822
studied previously \citep{OHalloran2010, Galametz2010}. However, the DGRs,
in the range 1/80--1/186, derived for these galaxies are higher.
The higher ratios may result from the assumption
of pure graphite or pure amorphous carbon dust, and from
the adopted gas and dust budget.

We note that the DGR of 1/190 was computed using the total derived
dust mass for NGC~4449, while the model covered the inner radius of 3.3~kpc,
the extent of the galaxy in the FIR,
and as such took into account only about 25 per~cent of the total observed gas
(Section~\ref{sec:assumptions_gas_density}).
If we considered only the gas present within the modelled region,
that would require all metals in NGC~4449 to be locked into dust. However, NGC~4449
is a highly dynamic system, postulated to have been involved in a merger
\citep{Theis2001} and possessing a recently discovered companion \citep{Rich2012}.
Since its optical body is not likely to be
an isolated system, it is informative to include 
the extended H\one\ envelope in the calculations, which yields a lower DGR of $\sim\;$1/760.
We also note that our gas masses do not include the mass of H$_2$
due to large uncertainties in the CO-to-H$_2$ conversion factors
for low metallicity galaxies. Inclusion of an H$_2$ mass of $\sim\;$$1.7\times10^8\;\textrm{M}_\odot$
could decrease the global DGR even further to $\sim\;$1/820.

On the other hand, an elevated flux at 850~$\mu$m, visible in Fig.~\ref{fig:n4449_mocassin_bestfitting}
and not matched by the model, may suggest the presence of
a significant mass of cold dust (e.g. \citealt{Galliano2003}).
Correctly accounting for the sub-mm emission is likely
to increase the total mass of dust, and consequently the DGR,
and can be crucial for accurate dust mass determinations (e.g. \citealt{Galametz2011}).

Finally, the dust mass predicted by our model can be
used to infer the fraction of metals locked into dust in NGC~4449.
Adopting the total H\one\ mass of $2.2\times10^9\;\textrm{M}_\odot$ (cf. \citealt{Swaters2002}, \citealt{Bajaja1994})
and given the total dust mass of $2.9\times10^6\;\textrm{M}_\odot$, the
carbon-to-silicate ratio of 1:3,
the LMC gas-phase C/O ratio of 0.5, the LMC gas-phase Si/O ratio
of 0.29 \citep{Russell1992} and the observed
mean oxygen abundance O/H of $1.95\times10^{-4}$ \citep{Vigroux1987},
we estimate that approximately 30 per~cent of carbon, 16 per~cent of oxygen
and 14 per~cent of silicon atoms
are locked in the solid phase as carbon and silicate dust.
These estimates are consistent with the efficiency of 0.12 implied for
SN 2003~gd \citep{Sugerman2006} and the efficiences expected in early-formed
galaxies \citep{Morgan2003}.
They are also broadly consistent with the dust condensation efficiency of 0.2
adopted independently for the theoretical supernova yields tabulated by \citet{Woosley1995} in order
to match the dust masses inferred for a number of supernovae (Section~\ref{sec:assumption_evolution_model}).

\input{./figures/figure.mocassin.variables.tex}
\input{./tables/table.MOCASSIN.SFH.tex}
\input{./tables/table.MOCASSIN.lines.tex}

\subsection{Spherical symmetry limitations} \label{sec:results_ionisation}

To achieve representative heating conditions in a spherical model,
the numerical setup must allow for enough ionizing photons to interact with
the correct mass of gas for
a given strength of the averaged ISRF. The images in
Fig.~\ref{fig:n4449_5imgs}, as well as the {\it Hubble} Space Telescope
observations of NGC~4449 presented by \citet{Annibali2008}
and the study of its young stellar clusters by \citet{Reines2008}
show a variety of distinct actively star-forming regions.
A spherically symmetric model must reproduce not only the
combined strength of the numerous individual sources, but
also the average ionization structure resulting from
the real three-dimensional distribution of these ionizing sources.
As a consequence, the required strength of the central ionizing source
may be slightly overestimated.

In general, Table~\ref{tab:MOCASSIN_lines} shows good agreement between the predicted
optical and infrared emission line fluxes and the observations. Since the
ISRF in our models falls with distance from the central source as $1/r^2$,
good agreement with observations
can only be achieved by modifying the
clumpiness or physical distribution of individual atomic or
molecular species.

For example, our results do not correctly predict the ionization structure of sulphur:
lines from the higher ionization potential
S$^{2+}$ ion are systematically underpredicted whereas lines from the
lower ionization potential S$^+$ ion are overpredicted.
This discrepancy
cannot be explained by the assumed global abundances. However, the predicted 
relative sulphur line fluxes shown in Table~\ref{tab:MOCASSIN_lines}
can be brought into very good agreement with observations by using a
filling factor of $\epsilon = 0.055$ for sulphur in place of the lower global
value of 0.033 adopted in this work.
Physically, a smoother ISM shortens the mean free path of
ionizing photons, increasing the degree of ionization within the modelled H~{\sc ii} region.
This result may suggest that
the nebular emission in NGC~4449 originates from two or more gas phases with
differing porosity, as well as from the transition zones between the phases.
We conclude that a multi-phase gas model has to be invoked
in our spherically symmetric approach in order to reproduce the details
of the observed ionization structure (cf. \citealt{Cormier2012}).

\subsection{Distributed ionizing sources} \label{sec:results_distributed}

A more realistic ISRF can be obtained by replacing the single
central ionizing source with a uniform distribution of sources
forming a central `ionizing sphere'. For this, we assumed a sphere
with $r=0.2\;\textrm{kpc}$ containing 100 identical sources,
each contributing \Lstar/100 to the total stellar luminosity.

Fig.~\ref{fig:n4449_mocassin_variables} (top panel,
black solid line) presents the predicted global SED
resulting from the model with distributed ionizing sources. The SED
is similar to the best-fitting model (red solid line),
except for the range $\sim\,$10--30~$\mu$m, where the
emission is weak in comparison. This discrepancy
in the strength of the thermal emission by warm dust can be explained
by the lower average photon energy available
to interact with the dust in the case of distributed sources:
dust in close proximity to one of
the ionizing sources emitting with \Lstar/100
is heated to much lower temperatures
than is a relatively smaller mass of dust placed near a single source
emitting with \Lstar.

The plots in Fig.~\ref{fig:n4449_mocassin_ionisation}
illustrate the corresponding differences in the ionization
structure as a function of radius.
For example, the overall ionization fraction of O$^{2+}$
is significantly higher in the model with distributed sources,
whereas the overall ionization fraction of O$^{3+}$
is significantly lower.
Since in this case the resulting integrated mass of O$^{3+}$ is smaller,
transitions of O$^{3+}$ are likely to produce
significantly lower line intensities. The opposite would be
expected for O$^{2+}$.
Indeed, in the distributed sources model shown in Fig.~\ref{fig:n4449_mocassin_ionisation},
the line fluxes predicted for the [O\three] lines were approximately 5 per~cent higher,
while the flux of the [O\two]$_{\lambda3727}$ line was 10 per~cent
lower than the corresponding fluxes in the single source model (Table~\ref{tab:MOCASSIN_lines}).
As discussed above, the key
physical factors determining individual line intensities
are the mean free paths to the ionizing source and
the strength of that ionizing source. Therefore,
in more realistic models with distributed sources,
the input luminosity \Lstar\ would be expected to be
lower by a few per~cent to maintain
good agreement between the predicted and observed line fluxes.

The physical distribution
of ionizing sources affects the global ionization structure
and, consequently, also affects the predicted line
intensities, derived elemental abundances (cf. \citealt{Ercolano2010})
and may affect the overall stellar luminosity \Lstar.
Therefore, we note that performing line fitting of
global emission lines is a valuable constraint on
the overall properties of a galaxy, but
may not be meaningful without
taking the morphology into consideration.

\input{./figures/figure.mocassin.ionisation.tex}

\section{Conclusions}\label{sec:conclusions}

In this paper, we used results from previous studies,
the available multiwavelength data, the spectral
fitting code STARLIGHT and a simple chemical
evolution code to construct a photoionization and
radiative transfer MOCASSIN model of NGC~4449
through an iterative scheme.

We assumed a simplified star formation history consisting of three continuous episodes.
Using the shape of the observed SED, the observed rate
of ionizing photons, $Q(\textrm{H}^0)$ derived from $L_{\mathrm{H}\alpha}$,
and the observed nebular emission line fluxes as criteria in our scheme,
we infer 3$\,\times\,$10$^6\;\textrm{M}_\odot$ for the youngest stellar population (0--10 Myr old),
which corresponds to an average recent star formation rate of 
0.28 $\textrm{M}_\odot\,\rm{yr}^{-1}$ or 0.42 $\textrm{M}_\odot\,\rm{yr}^{-1}$, assuming a
Kroupa IMF or a Salpeter IMF, respectively.
We infer a bolometric stellar luminosity of 5.7$\,\times\,$10$^9\;\textrm{L}_\odot$
generated by a total stellar mass of $\approx\;$$1\times10^9\;\textrm{M}_\odot$.
Although the ages of the oldest stellar populations
in NGC~4449 are not well constrained by the model, we note that
a very early onset, 12 Gyr ago or earlier, is selected by an independent
analysis using the spectral fitting code STARLIGHT. A more in-depth
study of stellar absorption line diagnostics at higher spectral resolutions
would be necessary to reach more definite conclusions about these older populations.

We modelled the entire stellar component self-consistently along with a mixture of
carbon and silicate dust. Our model yields a dust mass of $2.9\pm0.5\times10^6\;\textrm{M}_\odot$,
which includes 2 per~cent of PAHs. This estimate of $M_\textrm{dust}$
could be lower, if a higher proportion of carbon-rich dust is assumed.
Overall, the results from our chemical evolution model for a galaxy continuously forming stars
are consistent with the carbon to silicate dust mass ratio of 1:3 inferred for the LMC \citep{Weingartner2001}.
Interestingly, the dust condensation efficiency emerging from our MOCASSIN model and from
the observed abundance of oxygen in NGC~4449 is comparable with the value of 0.2 adopted independently
in our chemical evolution model to match the dust masses inferred for a range of supernovae.

We note that a DGR of 1/190 was derived for the modelled region with a radius of 3.3~kpc.
Including the extended H\one\ envelope and the molecular gas is likely to lower
the DGR to $\sim\;$1/800.
Similarly, our model does not account for cold dust, which may be responsible
for an excess at 850~$\mu$m. Taking this additional component into account may have
a significant effect on the derived DGR.

We conclude that our iterative scheme is a new tool, which can be
used to model both the dominant stellar populations
and the dust content in a self-consistent way. Although significant degeneracies in
the derived parameters are expected, they can be reduced by
supplementary spectroscopic data.

In the case of an irregular object, such as NGC~4449,
we note that the assumption of spherical symmetry
may lead to a misrepresentation
of the ISRF giving rise to unphysical conditions
near the central ionizing source. In particular,
our results have shown that a single ISM phase does not fully
reproduce the observed ionization structure, while
the assumption of a single DGR overpredicts the
mass of warm dust, suggesting that the radial density
profiles of dust and gas are significantly different.

Our scheme is easily expandable to three dimensions,
and in future work a realistic distribution of ionizing clusters
could be used together with a diffuse evolved stellar
population component and several gas and dust phases
to construct more detailed representations of galaxies.
In \citetalias{Karczewski2013_OBSERVATIONS} we present
a map of the recent star formation rate derived from
[Ne\two]$_{12.8}\,+\,$[Ne\three]$_{15.6}$, as well as spatially-resolved
measurements of the densities of ionized and neutral ISM phases,
which may be useful in constructing a three-dimensional model of NGC~4449
in the future.

\section{Acknowledgements}

O\L K acknowledges the support of UCL's Institute of Origins.
The authors acknowledge the use of the UCL {\it Legion}
High Performance Computing Facility, and associated support
services, in the completion of this work.

PACS has been developed by a consortium of institutes led by MPE (Germany) and including UVIE (Austria); KU Leuven, CSL, IMEC (Belgium); CEA, LAM (France); MPIA (Germany); INAF-IFSI/OAA/OAP/OAT, LENS, SISSA (Italy); IAC (Spain). This development has been supported by the funding agencies BMVIT (Austria), ESA-PRODEX (Belgium), CEA/CNES (France), DLR (Germany), ASI/INAF (Italy) and CICYT/MCYT (Spain).

SPIRE has been developed by a consortium of institutes led by Cardiff University (UK) and including University of Lethbridge (Canada); NAOC (China); CEA, OAMP (France); IFSI, University of Padua (Italy); IAC (Spain); Stockholm Observatory (Sweden); Imperial College London, RAL, UCL-MSSL, UKATC, University of Sussex (UK) and Caltech/JPL, IPAC, University of Colorado (USA). This development has been supported by national funding agencies: CSA (Canada); NAOC (China); CEA, CNES, CNRS (France); ASI (Italy); MCINN (Spain); Stockholm Observatory (Sweden); STFC, UKSA (UK) and NASA (USA).

\bibliographystyle{mn2e}
\bibliography{references.bib}

\appendix

\section{Photometry of extended sources with \Swift/UVOT}

The Ultraviolet/Optical Telescope (UVOT; \citealt{Swift_UVOT})
is one of three instruments on board \Swift\ \citep{Swift}
designed to detect and observe gamma-ray bursts
and their afterglows
in seven optical and ultraviolet bands. To this end,
the instrument is sensitive to single-photon events
and the data reduction software has been
tailored to point-source observations.
However, with its FOV of 17\arcmin$\,\times\,$17\arcmin,
the UVOT can also be of interest for studies of extended sources.
Although this FOV is considerably smaller to that of
\GALEX\ \citep{GALEX}, the UVOT offers a
higher angular resolution, with PSFs of only 2\farcsecp4--2\farcsecp9
for the three UV bands ({\it uvw2}, {\it uvm2} and {\it uvw1})
covering a comparable wavelength range to one of the two
\GALEX\ bands (Fig.~\ref{fig:Swift_UVOT};
\citealt{Morrissey2005,Breeveld2010}).

Obtaining reliable photometric measurements
for extended sources poses a technical challenge,
which is inherent to the design of the UVOT.
The signal from each incoming photon is electronically
multiplied to generate a splash of photons at the CCD stage
of the UVOT detector,
and the centroid of the splash gives positional information
to a sub-CCD-pixel accuracy.
However, the photon-counting detector has a readout rate
of $\sim\;$$90\;\mathrm{s}^{-1}$, which leads to systematic
undercounting (`coincidence loss' or `pile-up')
when two or more photons arrive in a similar location
on the detector within one frame. In such case,
the photons are not only counted as one, but the detection
is also misplaced by the centroiding algorithm.
At count rates of $\sim\;$10~s$^{-1}$ this effect
results in a 10 per~cent loss in the number of counts \citep{Poole2008}.

High-background cases of point-source observations
can be viewed as close analogues of extended sources.
A range of background levels was investigated in the models of
\citet{Breeveld2010}, who showed that backgrounds
higher than $\sim\,$0.07 s$^{-1}$ per unbinned image pixel
can no longer be fully corrected
for coincidence loss without introducing
an additional linear correction factor at each affected pixel.
This limiting
count rate was used by \citet{Hoversten2011}
in their analysis of M81 to highlight regions for
which coincidence loss is significant. Those regions
in M81 coincided with point-like star-forming knots and could
be corrected individually as point sources in the low background regime.
However, NGC~4449 is a dwarf starburst galaxy,
where the star formation activity, comparable to
that of M81 \citep{Gordon2004}, is localised in a relatively small volume of space.
The UV count rates in NGC~4449 are high throughout
the galaxy, as illustrated in Fig.~\ref{fig:Swift_uvw2},
making a similar approach not feasible.

\input{./figures/figure.Swift.UVOT.tex}
\input{./figures/figure.Swift.uvw2.tex}

Any method of obtaining photometry
must take into account the variation
of emission across the UVOT image, as areas
with higher count rates suffer from a
greater coincidence loss and require
a greater correction. However, coincidence loss
cannot be corrected for on a pixel-by-pixel basis because
the counts in neighbouring pixels are not independent of each other.
The position of each count in the image is
calculated from a photon splash over five or more
physical CCD pixels. Consequently,
all counts within the same CCD pixel (64 image pixels)
have been detected through similar photon splashes over
the same area of the CCD.
In the coincidence loss regime individual detections
in a particular area on the detector are not independent of each other
and should be considered collectively for coincidence loss correction.

\subsection{Isophotal correction}\label{appendix:isophotal}

We used NASA's HEAsoft (v6.7; released on 2009-08-19) to obtain
corrected photometry for three UV and three optical bands in NGC~4449.
The {\it white} band was excluded from this study.
The task {\sc uvotsource} was run repeatedly with two
user-defined apertures: one enclosing a region of interest
(or `source') and one defining the background.

The aperture size recommended for UVOT photometry is 5$^{\prime\prime}$
\citep{Poole2008}. By default, if the source aperture is greater than 5$^{\prime\prime}$, the
coincidence loss correction factor is determined from a 5$^{\prime\prime}$
circular region at the centre of the user-defined aperture.
Consequently, for larger apertures
the default procedure is likely to give a significantly biased corrected count rate,
depending on features present in the central area of the user-defined aperture.
Therefore, it is expected that {\sc uvotsource}
is generally not applicable `as-is' to sources
of angular extent greater than the size of the standard 5$^{\prime\prime}$ aperture.

In the method presented below, the image was divided into
regions following suitably chosen isophotal contours.
The coincidence loss correction
factors were determined from representative 5$^{\prime\prime}$ `test'
apertures and applied to correct photon count rates in the corresponding
regions. Afterwards, the background was subtracted and the count rates
were summed to give the total count rate.
An illustration of this isophotal setup is given in Fig.~\ref{fig:Swift_uvw2_regions}.
Region~1 in Fig.~\ref{fig:Swift_uvw2_regions} encloses
areas with count rates above the threshold of 0.028 s$^{-1}$ per pixel
for $2\times2$ binning, where the effects of coincidence loss become non-negligible \citep{Breeveld2010,Hoversten2011}.
Regions 2--4 in Fig.~\ref{fig:Swift_uvw2_regions} enclose areas with count rates of 0.1, 0.2 and 0.3 s$^{-1}$ per pixel.
Finally, region~0 is an ellipse enclosing the entire galaxy but not aligned with its optical centre.

\input{./figures/figure.Swift.uvw2.regions.tex}

Measurements in all individual regions were performed by running the HEAsoft command:

\begin{small}
\begin{verbatim}
uvotsource image=image_sk_coadd.fits \
           expfile=image_ex_coadd.fits \
           srcreg=REGION.reg bkgreg=bkg.reg \
           apercorr=NONE centroid=NO clobber=NO \
           frametime=0.0110329 output=ALL sigma=5 \
           syserr=YES outfile=photometry.fits
\end{verbatim}
\end{small}
where {\tt REGION.reg} is a description of
{\it (i)} a test region 0--4, or {\it (ii)}
one of the regions comprising regions 0--4
in Fig.~\ref{fig:Swift_uvw2_regions}.
It is useful to define $A$, $C_{\mathrm{r(,err)}}$,
$C_{\mathrm{test(,err)}}$, $f_{\mathrm{test}}$ and
$C_{\mathrm{bkg(,err)}}$ based on the following
columns in the output {\sc fits} table:

\begin{center}\begin{tabular}{l p{4.8cm}}
\texttt{SRC\_AREA} & area of the source aperture in arcsec$^2$, $A$ \smallskip \tabularnewline
\texttt{RAW\_TOT\_RATE(\_ERR)} & raw count rate (error) in the source aperture, $C_{\mathrm{r(,err)}}$ \smallskip  \tabularnewline
\texttt{COI\_TOT\_RATE(\_ERR)} & corrected count rate (error) in the source aperture, $C_{\mathrm{test(,err)}}$ \smallskip \tabularnewline
\texttt{COI\_STD\_FACTOR} & coincidence loss correction factor in the source aperture, $f_{\mathrm{test}}$ \smallskip \tabularnewline
\texttt{COI\_BKG\_RATE(\_ERR)} & corrected count rate (error) per arcsec$^2$ in the background aperture, $C_{\mathrm{bkg(,err)}}$. \tabularnewline
\end{tabular}\end{center}

The values of $C_{\mathrm{test(,err)}}$ and $f_{\mathrm{test}}$ were measured
from the five test regions. The measurements from individual regions in each of
regions 0--4 were combined to obtain integrated measurements for regions 0--4,
$A_n$, $C_{\mathrm{r},n}$ and $C_{\mathrm{r,err},n}$. Since the same background aperture {\tt bkg.reg},
defined as an elliptical annulus, was used in all instances, 
$C_{\mathrm{bkg(,err)}}$ represents a single measurement.
The total corrected and background-subtracted count rate $C_{\mathrm{src}}$ was then
obtained by summing individually corrected stripes, or `isophotes':

\begin{equation} \label{eq:Swift_countrate}
C_{\mathrm{src}} = \sum_{n=0}^{4} \; C_{\mathrm{src},n}, \end{equation}
where
\begin{equation}
C_{\mathrm{src},n}=C_{\mathrm{i},n}\times f_{\mathrm{test},n} -C_{\mathrm{bkg}}\times A_{\mathrm{i},n}, \end{equation}

\begin{equation}
C_{\mathrm{i},n}=\begin{cases}
C_{\mathrm{r},n}-C_{\mathrm{r},n+1} & 0 \leq n \leq 3 \\
C_{\mathrm{r},4} & n = 4,\end{cases} \end{equation}
and
\begin{equation}
A_{\mathrm{i},n}=\begin{cases}
A_n-A_{n+1} & 0 \leq n \leq 3 \\
A_4 & n = 4.\end{cases} \end{equation}

The upper limit to the statistical error in $C_{\mathrm{src},n}$ can be estimated
by assuming that the fractional error in the count rate $C_{\mathrm{i},n}$ in
each isophotal region is equal to the fractional error in the corrected count rates in the
corresponding test region given by $C_{\mathrm{test,err},n}/C_{\mathrm{test},n}$.
The errors associated with the corrected count rates for higher correction factors
are significantly higher than the Poisson error, as shown by \citet{Kuin2008}.
Both $C_{\mathrm{test,err}}$ and $C_{\mathrm{bkg,err}}$
are binomial errors associated with the corrected count rates \citep{Poole2008,Kuin2008}.

\subsection{Results}

Detailed measurements in band {\it uvw2} are presented in
Table~\ref{tab:swift_photometry_uvw2}. These results show that
even in region~1 the average count rate per
pixel is 0.055 s$^{-1}$, which is significantly higher than
the coincidence loss threshold of 0.028 s$^{-1}$ \citep{Hoversten2011}.

The total count rates in all six UVOT bands are
given in Table~\ref{tab:swift_photometry_summary}.
The corresponding fluxes were obtained
using the calibrations of \citet{Breeveld2010_AB},
and were subsequently corrected for foreground
extinction using $E(B-V) = 0.019$ \citep{Schlegel1998}
and the extinction law of \citet{CCM} with $R_V=3.1$.

The uncertainties in the count rates 
listed in Table~\ref{tab:swift_photometry_summary}
combine {\it (i)} the binomial errors in each of the
test regions 0--4 \citep{Kuin2008} scaled to the area
of the corresponding region, {\it (ii)} the estimated uncertainties due
to high background \citep{Breeveld2010}
and {\it (iii)} the uncertainty resulting from
the choice of thresholds for the five regions.
For the UV and optical bands,
the uncertainties in {\it (ii)}
were estimated at 6~and 8 per~cent, respectively.
The uncertainties in {\it (iii)} were found to
be at most 2 per~cent in all bands.
Thus, the statistical and systematic uncertainties in the
final fluxes (Table~\ref{tab:swift_photometry_summary})
amount to $\sim\;$7 and $\sim\;$9 per~cent overall for the UV and
the optical bands.

\input{./tables/table.appendix.uvw2.tex}
\input{./tables/table.appendix.photometry.tex}

Table~\ref{tab:swift_photometry_summary} also
gives total fluxes calculated
by {\sc uvotsource} directly from region~0 in the `as-is' approach,
i.e., ignoring the spatial variations in the correction factors
across UVOT images. In bands {\it uvw2}, {\it uvm2} and {\it uvw1}
the fluxes agree to within 7~per cent, whereas in the optical
bands the discrepancy is more significant.
The agreement in the UV bands may suggest that the centre
of region~0 probed the global average of the count rate distribution for the UV bands,
weighted by the extent of coincidence loss.
However, in sources that are generally fainter in the UV,
better agreement in these bands may
be expected as the required correction factors are smaller.
Although the values of $f_{\mathrm{max}}$ listed in Table~\ref{tab:swift_photometry_summary}
show that the maximum correction factors were high in all bands, the
emission in the UV is sharply peaked compared to the more uniform emission in the optical.
This suggests that applying corrections to the UV bands is likely to be
affected by a smaller systematic uncertainty than applying similar corrections
to the optical bands. In general, these results confirm that {\sc uvotsource} is
not applicable `as-is' to
photometric measurements of extended sources, as expected from Section~\ref{appendix:isophotal}.

The UV and optical fluxes in Table~\ref{tab:photometry_table}
are consistent with those obtained from \GALEX\ and 
SDSS. A plot of the global SED of NGC~4449 in the range 1000--7000 \r{A} is
presented in Fig.~\ref{fig:Swift_SED}.

In the UV, the fluxes derived from \GALEX\ and \Swift/UVOT agree
very well, but the absolute NUV flux is slightly lower
than the corresponding fluxes obtained from \Swift/UVOT.
It should be noted that \GALEX, in a similar way to \Swift/UVOT,
also suffers from local non-linearities near bright
sources. These are more difficult to quantify
and are not automatically corrected for \citep{Morrissey2007}.
The count rates measured for NGC~4449 in
a 3\arcmin\ diameter aperture ($\sim\;$1000~s$^{-1}$ in FUV and
$\sim\;$4000~s$^{-1}$ in NUV) suggest that this effect is likely to 
lower the measured count rates and 
add a significant systematic uncertainty to the derived \GALEX\ fluxes.
Therefore, the true \GALEX\ FUV and NUV fluxes 
may be higher and may be associated with larger uncertainties than those
shown in Fig.~\ref{fig:Swift_SED}.

\subsection{Conclusions}

In the method presented above we used
the existing understanding of coincidence loss for
point-source observations and applied it to 
obtain photometric measurements for an extended source,
NGC~4449, in three UV and three optical \Swift/UVOT bands. 
The derived fluxes are in good agreement with the
available global \GALEX\ and SDSS photometry, with
the overall uncertainties estimated at 7~and 9 per~cent for the
UV and the optical bands. 

Extended-source observations with the three narrow UV bands
and the small PSFs of \Swift/UVOT enable more detailed
studies of the youngest stellar component in galaxies.
Although the uncertainties
in our method are comparable with those of
\GALEX, we note that at higher count rates
\GALEX\ also suffers from non-linearity,
which is not corrected for or included in
the published uncertainties.

\input{./figures/figure.Swift.SED.tex}

\bsp

\label{lastpage}

\end{document}

%% file: figures/figure.n4449.5images.tex
\begin{figure*}
\centering
\begin{minipage}[t]{160mm}
\includegraphics[scale=1]{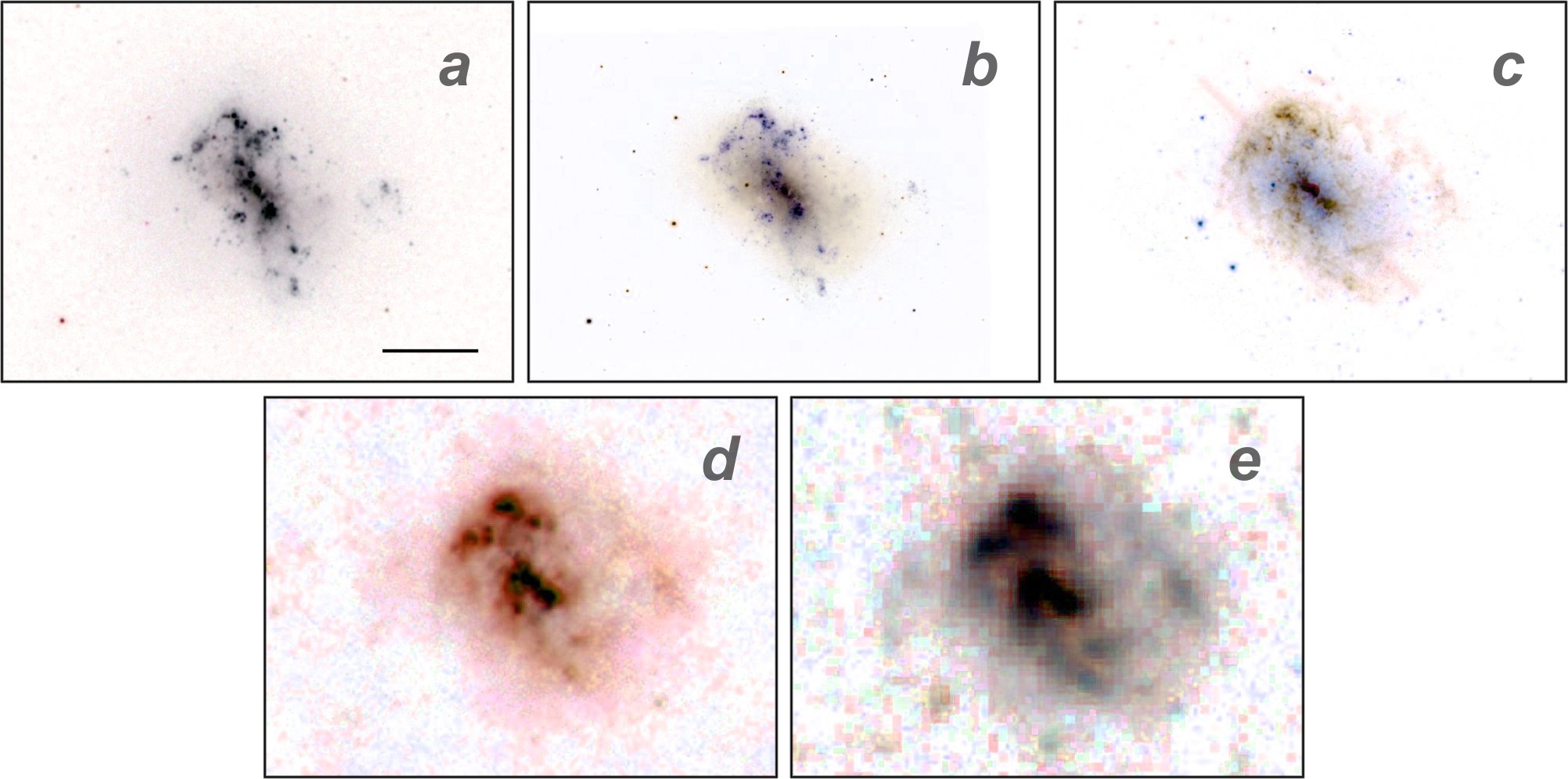}
\caption{False-colour (R, G, B) broadband images of NGC~4449: \textit{(a)} \Swift/UVOT (2486~\AA, 2221~\AA, 1991~\AA), \textit{(b)} SDSS (\textit{r}, \textit{g}, \textit{u}), \textit{(c)} \Spitzer/IRAC (8~$\mu$m, 5.8~$\mu$m, 3.6~$\mu$m), \textit{(d)} \Herschel/PACS (160~$\mu$m, 100~$\mu$m, 70~$\mu$m) and \textit{(e)} \Herschel/SPIRE (500~$\mu$m, 350~$\mu$m, 250~$\mu$m). North is up, east is to the left. The images are centred at 12$^{\textrm{h}}$28$^{\textrm{m}}$11\farcsec1, +44$^{\circ}$05\arcmin37\arcsec\ (J2000). The bar in the top left image is 2\arcmin\ in length (2.2~kpc).\label{fig:n4449_5imgs}}
\end{minipage}
\end{figure*}

%% file: tables/table.observations1.tex
\begin{table*}
\begin{minipage}[t]{120mm}
\begin{center}
\begin{tabular}{llrrr}

\hline
Instrument & Bandpass/$\lambda_\mathrm{eff}$ & Time (s) & Date observed & Observation ID \tabularnewline \hline \hline \noalign{\smallskip}

\multicolumn{5}{c}{New observations} \tabularnewline \noalign{\smallskip}

\Herschel/PACS & 70~$\mu$m + 160~$\mu$m & 3497 & 2011 May 16 & 1342221125 \tabularnewline
& 70~$\mu$m + 160~$\mu$m & 3497 & 2011 May 16 & 1342221126 \tabularnewline
& 100~$\mu$m + 160~$\mu$m & 3497 & 2011 May 16 & 1342221127 \tabularnewline
& 100~$\mu$m + 160~$\mu$m & 3497 & 2011 May 16 & 1342221128 \tabularnewline \noalign{\smallskip}

\Herschel/SPIRE & 250~$\mu$m, 350~$\mu$m, 500~$\mu$m & 1035 & 2010 June 12 & 1342198243 \tabularnewline \noalign{\medskip}

\multicolumn{5}{c}{Archival observations} \tabularnewline \noalign{\smallskip}

\Swift/UVOT & \textit{uvw2}/2030\AA & 1544 & 2007 Mar 27 & 00035873001 \tabularnewline
& \textit{uvm2}/2231\AA & 1091 & 2007 Mar 27 & 00035873001 \tabularnewline
& \textit{uvw1}/2634\AA & 771 & 2007 Mar 27 & 00035873001 \tabularnewline
& \textit{u}/3501\AA & 385 & 2007 Mar 27 & 00035873001 \tabularnewline
& \textit{b}/4329\AA & 385 & 2007 Mar 27 & 00035873001 \tabularnewline
& \textit{v}/5402\AA & 385 & 2007 Mar 27 & 00035873001 \tabularnewline \noalign{\smallskip}

\GALEX & FUV/1539\AA & 835 & 2006 Mar 15 & tile 5228 \tabularnewline
               & NUV/2316\AA & 765 & 2006 Mar 15 & tile 5228 \tabularnewline \noalign{\smallskip}

SDSS & \textit{u}/3551\AA & 54 & 2003 Mar 24 & 3813/1/41/237/241 \tabularnewline
& \textit{g}/4686\AA & 54 & 2003 Mar 24 & 3813/1/41/237/245 \tabularnewline
& \textit{r}/6165\AA & 54 & 2003 Mar 24 & 3813/1/41/237/237 \tabularnewline
& \textit{i}/7481\AA & 54 & 2003 Mar 24 & 3813/1/41/237/239 \tabularnewline
& \textit{z}/8931\AA & 54 & 2003 Mar 24 & 3813/1/41/237/243 \tabularnewline \noalign{\smallskip}

\hline

\end{tabular}
\caption{List of observations.\label{tab:observations_table}}
\end{center}
\end{minipage}
\end{table*}

%% file: tables/table.photometry.tex
\begin{table*}
\centering
\begin{minipage}[t]{105mm}
\begin{center}
\begin{tabular}{@{\hspace{3pt}}lll@{\hspace{6pt}}cc@{\hspace{3pt}}}

\hline
Survey/instrument & $\lambda_\mathrm{eff}$ & $F_\nu$ & Aperture$^a$ & Reference \tabularnewline \hline \hline

\GALEX & 1539 \AA & 152 $\pm$ 12 mJy & 11\farcmin5 & (1)\tabularnewline
& 2316 \AA & 183 $\pm$ 9 mJy & 7\farcmin5 & \tabularnewline \noalign{\smallskip}

\Swift/UVOT & 1991 \AA & 175 $\pm$ 12 mJy & 8\farcmin5$\;\times\;$6\arcmin & (1)\tabularnewline
& 2221 \AA & 189 $\pm$ 14 mJy & \tabularnewline
& 2486 \AA & 202 $\pm$ 15 mJy & \tabularnewline
& 3442 \AA & 249 $\pm$ 22 mJy & \tabularnewline
& 4321 \AA & 463 $\pm$ 41 mJy & \tabularnewline
& 5410 \AA & 611 $\pm$ 57 mJy & \tabularnewline \noalign{\smallskip}

SDSS & 3551 \AA & 242 $\pm$ 5 mJy & 6\farcmin5$\;\times\;$4\farcmin7 & (1) \tabularnewline
& 4686 \AA & 457 $\pm$ 14 mJy & \tabularnewline
& 6165 \AA & 573 $\pm$ 40 mJy & \tabularnewline
& 7481 \AA & 622 $\pm$ 50 mJy & \tabularnewline
& 8931 \AA & 683 $\pm$ 34 mJy & \tabularnewline \noalign{\smallskip}

2MASS & 1.24 $\mu$m & 916 $\pm$ 21 mJy & 8\arcmin$^b$ & (2) \tabularnewline
& 1.66 $\mu$m & 1070 $\pm$ 30 mJy & \tabularnewline
& 2.16 $\mu$m & 839 $\pm$ 31 mJy & \tabularnewline \noalign{\smallskip}

\WISE & 3.4 $\mu$m & 391 $\pm$ 7 mJy & 4\farcmin0$\;\times\;$2\farcmin8 & (3) \tabularnewline
& 4.6 $\mu$m & 251 $\pm$ 4 mJy & & \tabularnewline
& 12 $\mu$m & 954 $\pm$ 17 mJy & & \tabularnewline
& 22 $\mu$m & 2860 $\pm$ 60 mJy & & \tabularnewline \noalign{\smallskip}

\Spitzer/IRAC & 3.6 $\mu$m & 493 $\pm$ 15 mJy & $\dots$ & (4) \tabularnewline
& 4.5 $\mu$m & 317 $\pm$ 10 mJy & & \tabularnewline
& 5.8 $\mu$m & 615 $\pm$ 19 mJy & & \tabularnewline
& 8 $\mu$m & 1420 $\pm$ 40 mJy & & \tabularnewline \noalign{\smallskip}

IRAS & 12 $\mu$m & 2.1 $\pm$ 0.2 Jy & 8\arcmin & (5) \tabularnewline
& 25 $\mu$m & 4.7 $\pm$ 0.5 Jy & & \tabularnewline
& 60 $\mu$m & 36 $\pm$ 3 Jy & & \tabularnewline \noalign{\smallskip}

\Spitzer/MIPS & 24 $\mu$m & 3.29 $\pm$ 0.13 Jy & 9\farcmin3$\;\times\;$6\farcmin6 & (6) \tabularnewline
& 70 $\mu$m & 43.8 $\pm$ 4.4 Jy & & \tabularnewline
& 160 $\mu$m & 78.1 $\pm$ 9.4 Jy & & \tabularnewline \noalign{\smallskip}

\Herschel/PACS & 70 $\mu$m & 49.3 $\pm$ 2.5 Jy & 8\farcmin3 & (7) \tabularnewline
& 100 $\mu$m & 75.9 $\pm$ 3.8 Jy & & \tabularnewline
& 160 $\mu$m & 79.5 $\pm$ 4.0 Jy & & \tabularnewline \noalign{\smallskip}

\Herschel/SPIRE & 250 $\mu$m & 35.7 $\pm$ 2.5 Jy & 8\farcmin3 & (7) \tabularnewline
& 350 $\mu$m & 16.2 $\pm$ 1.1 Jy & & \tabularnewline
& 500 $\mu$m & 5.6 $\pm$ 0.4 Jy & & \tabularnewline \noalign{\smallskip}

\Planck & 350 $\mu$m & 16.14 $\pm$ 0.34 Jy & 6\farcmin1$\;\times\;$5\farcmin2$^c$ & (8) \tabularnewline
& 550 $\mu$m & 4.97 $\pm$ 0.24 Jy & 5\farcmin8$\;\times\;$4\farcmin8$^c$ & \tabularnewline
& 850 $\mu$m & 1.45 $\pm$ 0.15 Jy & 6\farcmin0$\;\times\;$4\farcmin4$^c$ & \tabularnewline \noalign{\smallskip}

IRAM & 1.2 mm & 260 $\pm$ 40 mJy & two fields 2\farcmin3 & (9) \tabularnewline \hline

\end{tabular}

\begin{center}
\begin{footnotesize}
\vspace{-2mm}
\parbox{100mm}{$^a$ given as diameters for circular apertures or major$\;\times\;$minor axes for elliptical apertures; $^b$ major axis of the `total' aperture; $^c$ FWHM of a Gaussian fit}
\end{footnotesize}
\end{center}

\caption{Summary of the derived photometric data for NGC 4449.
All fluxes have been corrected for the foreground
extinction using $E(B-V) = 0.019$ \citep{Schlegel1998}
and the extinction law of \citet{CCM} with $R_V=3.1$. References: (1) this work; (2) 2MASS All-Sky Extended Source Catalog \citep{2MASS}; (3) \WISE\ All-Sky Source Catalog \citep{WISE}; (4) \citet{Engelbracht2008}; (5) \citet{Hunter1986}; (6) \citet{Bendo2012b}; (7) \citet{Remy2013}; (8) \Planck\ Early Release Compact Source Catalogue \citep{Planck_ERCSC}; (9) \citet{Bottner2003}.\label{tab:photometry_table}}
\end{center}
\end{minipage}
\end{table*}

%% file: figures/figure.mocassin.method.tex
\begin{figure}
\centering
\begin{minipage}[t]{80mm}
\includegraphics[width=80mm]{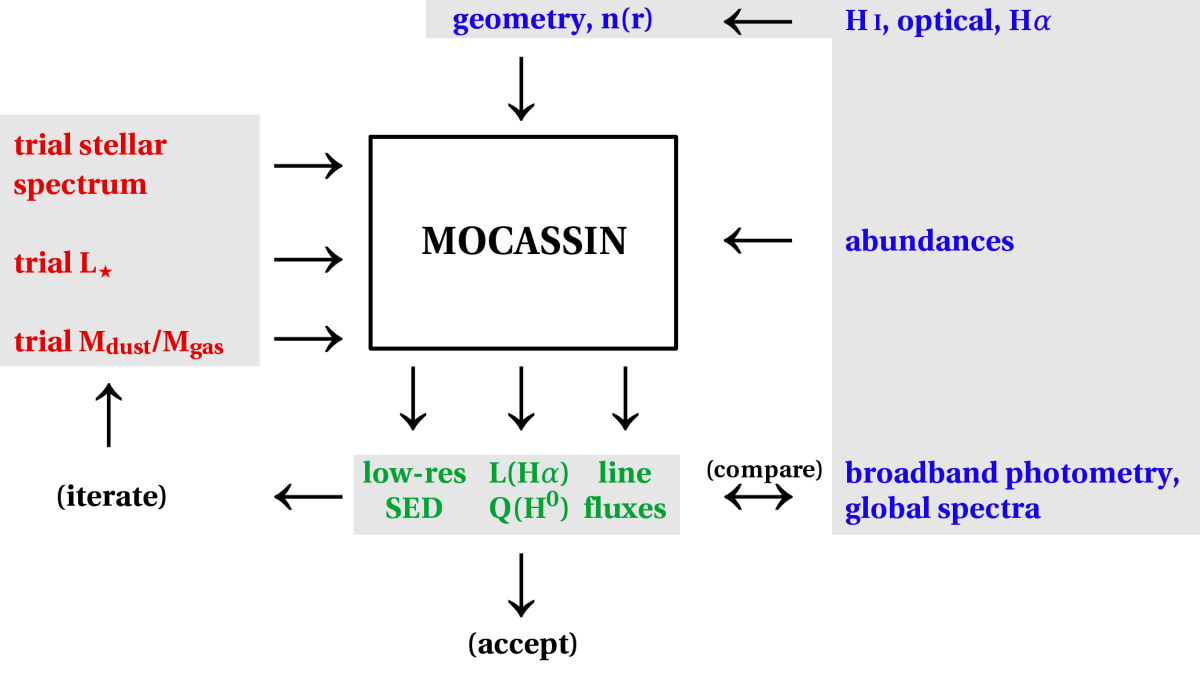}
\vspace{-6mm}
\caption{Summary of the modelling method. The shaded regions group
together the input variables (left) the outputs (middle) and the observational constrains
(right).\label{fig:n4449_mocassin_method}}
\end{minipage}
\end{figure}

%% file: figures/figure.STARLIGHT.example.tex
\begin{figure}
\centering
\begin{minipage}[t]{80mm}
\includegraphics[width=80mm]{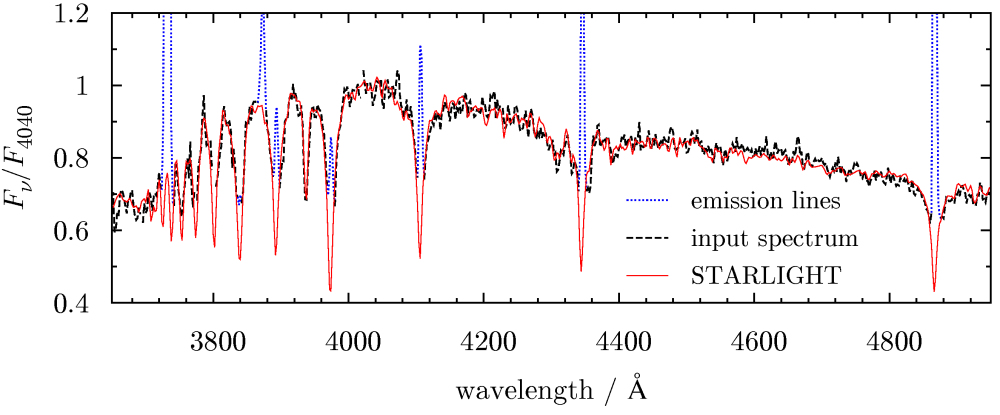}
\vspace{-3mm}
\caption{One of 12 STARLIGHT fits, in the range 3700--4900~\AA, to the global spectrum of NGC~4449 acquired
by \citet{Kennicutt1992}. The input spectrum (black line)
was decomposed using 33 different stellar ages at $Z/\textrm{Z}_\odot=0.2$.
Contaminating emission lines in the observed spectrum, shown in blue, were masked out.
See text for more details.\label{fig:n4449_STARLIGHT_example}}
\end{minipage}
\end{figure}

%% file: tables/table.STARLIGHT.results.tex
\begin{table}
\centering
\begin{minipage}[t]{80mm}
\begin{center}
\begin{tabular}{@{\hspace{3pt}}c@{\hspace{12pt}}ccc@{\hspace{3pt}}}
\hline \noalign{\smallskip}
& \multicolumn{3}{c}{Mass fraction by age of stellar population} \tabularnewline \noalign{\smallskip} \cline{2-4} \noalign{\smallskip}
Metallicity & Young & Intermediate & Old \tabularnewline
& ($\lesssim\,$10 Myr) & ($\sim\,$10$^2$ Myr) & ($\gtrsim\,$10$^3$ Myr) \tabularnewline
\hline
\hline \noalign{\smallskip}	
$0.2\textrm{Z}_\odot$ & 1\% & 20--25\% & 60--75\% \medskip \tabularnewline 
$0.4\textrm{Z}_\odot$ & $<\,$2\% & 2\% & \medskip \tabularnewline
$\textrm{Z}_\odot$ & $<\,$1\% & $<\,$7\% & \smallskip \tabularnewline
\hline
\end{tabular}

\end{center}
\vspace{-1mm}

\caption{Summary of results from STARLIGHT fits to the global spectrum of NGC~4449 acquired by \citet{Kennicutt1992}.
The figures show the percentages of the total mass of stars contributed by populations in the given age and metallicity bin.\label{tab:starlight}}
\end{minipage}
\end{table}

%% file: figures/figure.n4449.dust.evolution.tex
\begin{figure*}
\centering
\begin{minipage}[t]{170mm}
\begin{center}
\includegraphics[width=170mm]{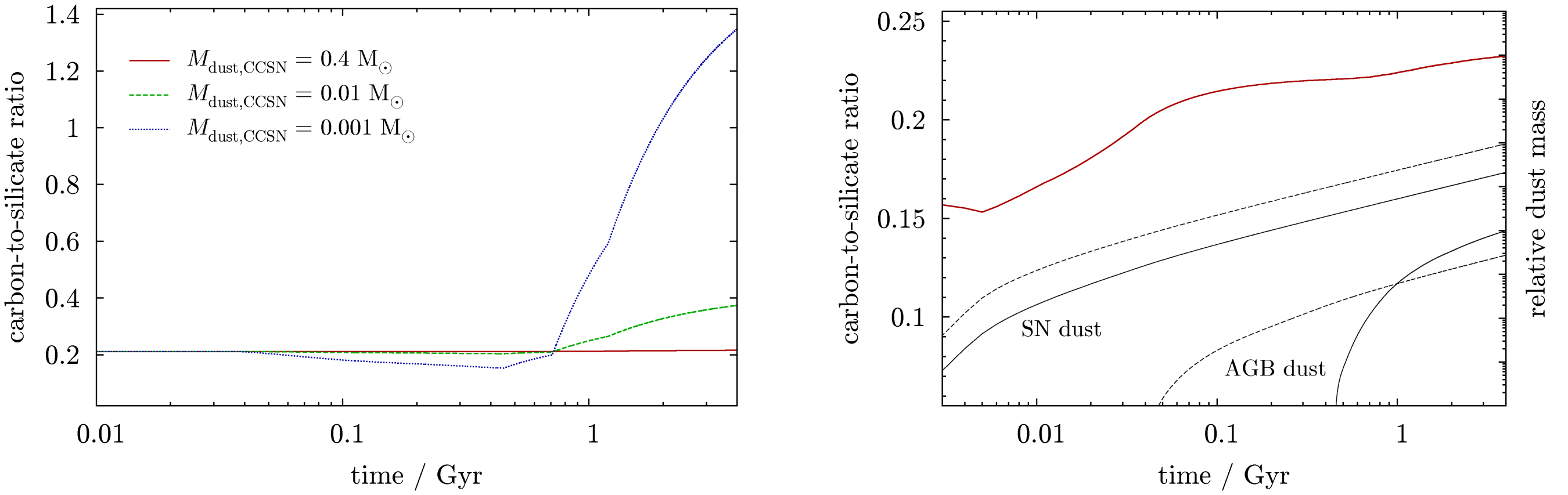}
\end{center}
\caption{A simple model of the evolution of the global relative carbon to silicate dust content of NGC~4449, assuming continuous star formation and a constant metallicity $Z/\textrm{Z}_\odot\sim0.1$. The left panel shows the evolution of the carbon-to-silicate ratio for supernova dust yields identical to those derived by \citet{Matsuura2011} for SN~1987A (solid line), adopting their lower limit of $M_\textrm{dust}=0.4\;\textrm{M}_\odot$ and a carbon-to-silicate ratio of 0.21 (their Model~1). Similar models with dust masses scaled to $0.01\;\textrm{M}_\odot$ (dashed line) and $0.001\;\textrm{M}_\odot$ (dotted line) are also presented, where the $0.001\;\textrm{M}_\odot$ model may be viewed as one where core-collapse supernovae are not important in global dust production.
The right panel shows the evolution of the carbon-to-silicate ratio for supernova dust yields based on the elemental yields of \citet[ solid line]{Woosley1995}. The thin solid and dashed lines show, respectively, the relative masses of carbon and silicate dust produced.\label{fig:n4449_dust_evolution_model}}
\end{minipage}
\end{figure*}

%% file: tables/table.MOCASSIN.inputs.tex
\begin{table*}
\centering
\begin{minipage}[t]{150mm}
\begin{center}
\begin{tabular}{p{3cm} p{5.5cm} p{5.5cm}}

\hline
Parameter & Value & Comments \tabularnewline \hline \hline \noalign{\medskip}

\multicolumn{3}{c}{\it{Numerical setup}} \tabularnewline \noalign{\smallskip}

geometry & 3D, spherically-symmetric & \tabularnewline
grid size & $80\times80\times80$ & ionizing source at $(x,y,z) = (0,0,0)$ \tabularnewline
number of photons & $2\times10^8$ & over all wavelengths\tabularnewline \noalign{\medskip}

\multicolumn{3}{c}{\it{Observationally constrained parameters}} \tabularnewline \noalign{\smallskip}

physical radius & 3.3 kpc & \tabularnewline
gas distribution & exponential-like & derived from the H~{\sc i} profile of \newline \citet{Swaters2002} \tabularnewline
$M_\textrm{gas}$ & $0.55\times10^9\;\textrm{M}_\odot$ & computed for $0\leq r \leq 3.3\;\textrm{kpc}$; molecular gas not included \tabularnewline
filling factor ($\epsilon$) & 0.033 & \tabularnewline
chemical elements & H, He, C, N, O, Ne, S & abundances constant throughout galaxy \tabularnewline
$Q(\textrm{H}^0)$ & $4.84 \times 10^{52}\;\textrm{s}^{-1}$ & based on $L_{\textrm{H}\alpha}$ \citep{Hunter1999} \tabularnewline \noalign{\medskip}

\multicolumn{3}{c}{\it{Adopted dust characteristics}} \tabularnewline \noalign{\smallskip}

dust composition & amorphous carbon and silicates (1:3) & \citet{Hanner1988}; \citet{Laor1993} \tabularnewline
dust grain sizes & 0.005--0.25 $\mu$m (20 sizes) & \citet{MRN77} \tabularnewline
PAH grain sizes & 3.5--30 \r{A} (10 sizes) & PAHs modelled separately; \newline \citet{Weingartner2001} \tabularnewline \noalign{\medskip}

\multicolumn{3}{c}{\it{Best-fitting parameters}} \tabularnewline \noalign{\smallskip}

$L_\star$ & $5.7\times10^9\,\textrm{L}_\odot$ & \tabularnewline
$M_\star$ & $1.05\pm0.15\times10^9\;\textrm{M}_\odot$ & \tabularnewline
$M_\textrm{dust}$ & $2.9\pm0.5\times10^6\;\textrm{M}_\odot$ & including PAH grains \tabularnewline
$M_\textrm{PAH}$ & $0.058\pm0.005\times10^6\;\textrm{M}_\odot$ & ionized PAH grains; \citet{Draine2007} \tabularnewline
\DGR & 1/190 & effective; \newline 1/1000 in centre, 1/182 otherwise \tabularnewline \noalign{\medskip}

\hline

\end{tabular}

\caption{Summary of the input parameters and the results from the best-fitting MOCASSIN model of NGC~4449.\label{tab:MOCASSIN_inputs}}
\end{center}
\end{minipage}
\end{table*}

%% file: figures/figure.mocassin.bestfitting.tex
\begin{figure*}
\centering
\begin{minipage}[t]{170mm}
\includegraphics[width=170mm]{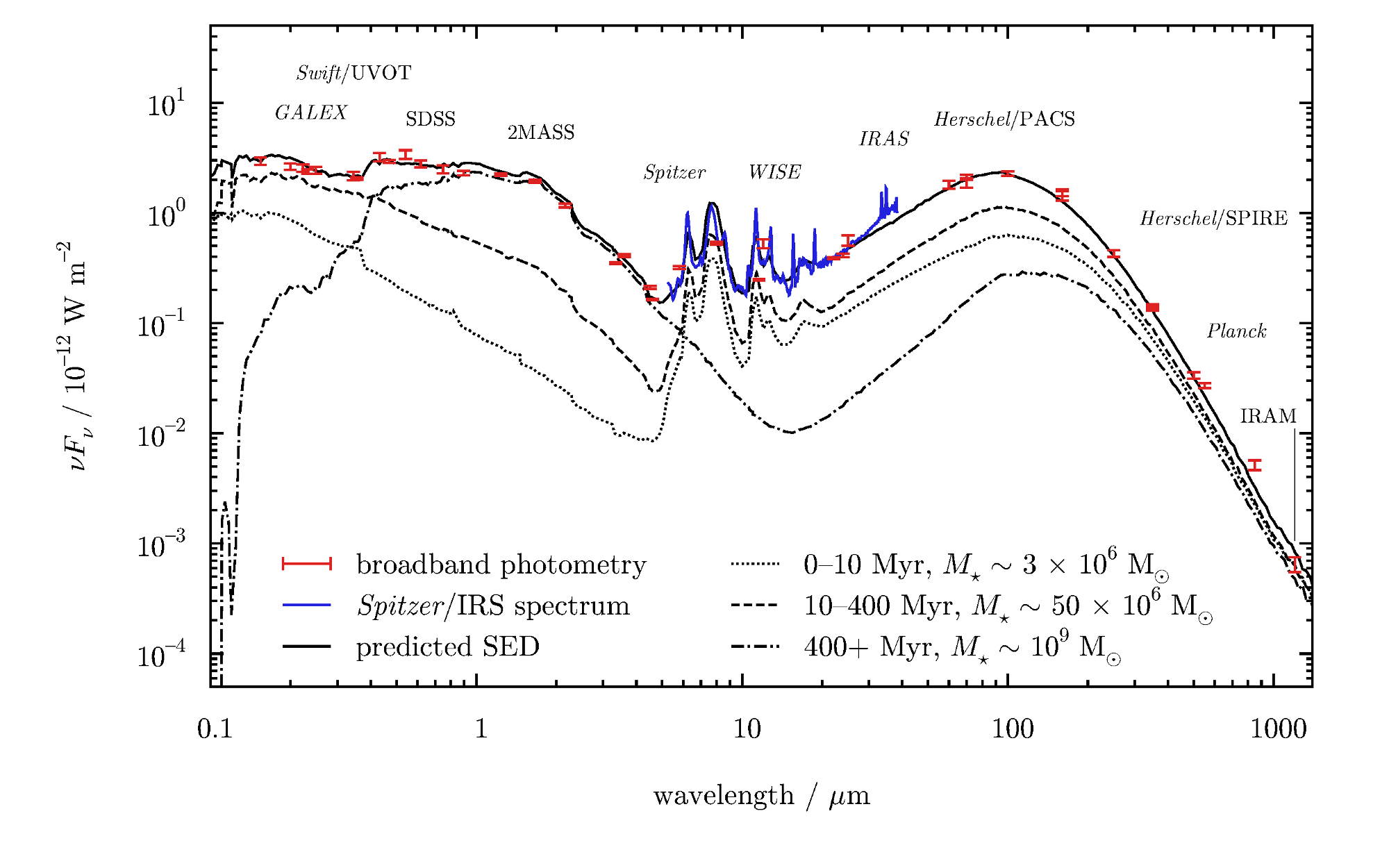}
\vspace{-8mm}
\caption{SED of the best-fitting MOCASSIN model of NGC~4449. The three star formation episodes were modelled simultaneously, but their individual SEDs are also shown for comparison. The photometric measurements have been corrected for foreground extinction, and the mid-IR spectrum acquired with \Spitzer/IRS is discussed in detail in \citetalias{Karczewski2013_OBSERVATIONS}.\label{fig:n4449_mocassin_bestfitting}}
\end{minipage}
\end{figure*}

%% file: figures/figure.mocassin.variables.tex
\begin{figure*}
\centering
\begin{minipage}[t]{170mm}
\includegraphics[width=170mm]{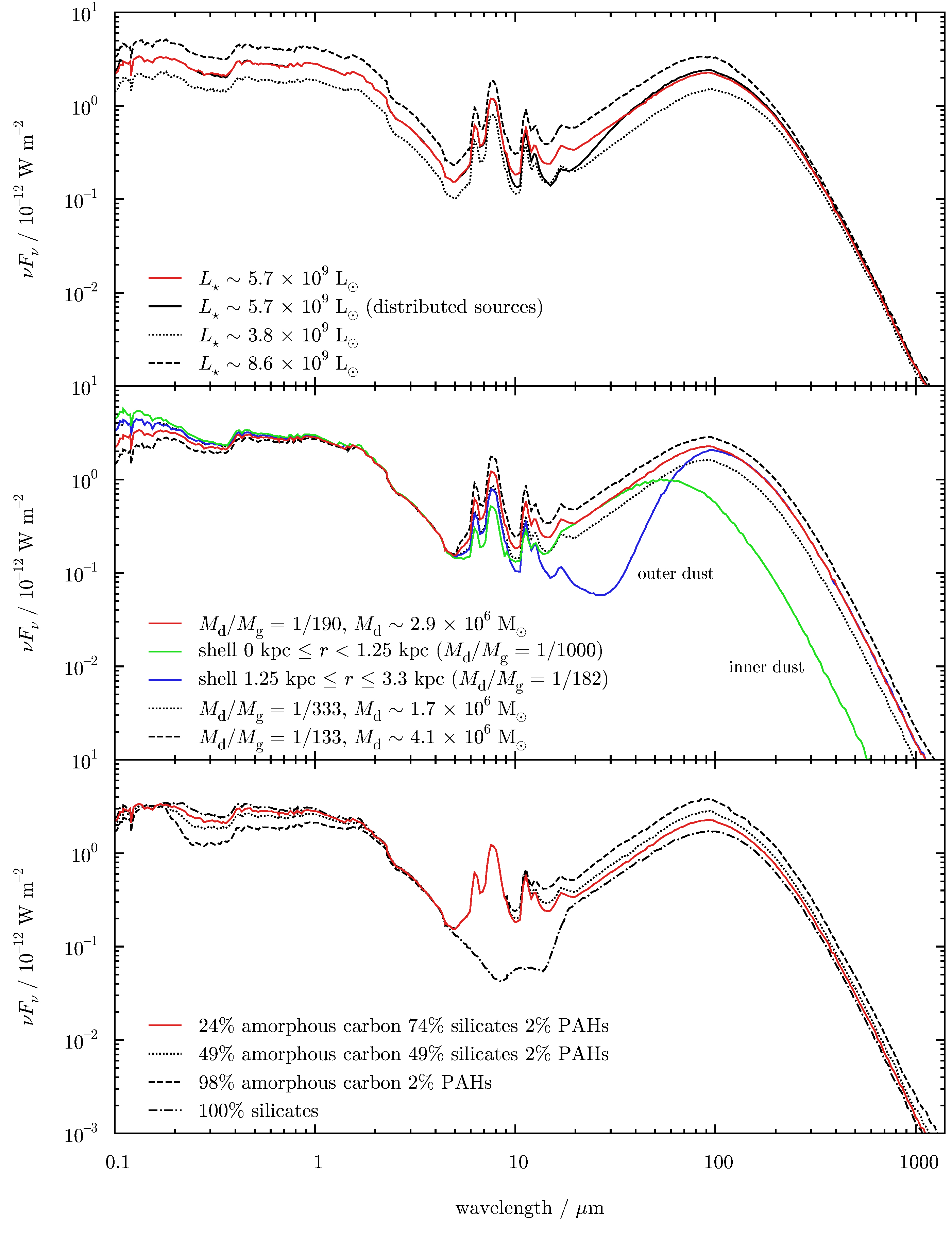}
\vspace{-4mm}
\caption{The effects of varying \Lstar\ (top panel), the DGR (middle panel) and the relative dust composition (bottom panel) on the predicted SED. While \Lstar, the DGR and the distribution of the DGR were free variables in the models, the dust composition was fixed (Section~\ref{sec:assumptions_dust_composition}) and its effect is presented here for illustration purposes only. The red solid lines represent the best-fitting model. See text for more details.\label{fig:n4449_mocassin_variables}}
\end{minipage}
\end{figure*}

%% file: tables/table.MOCASSIN.SFH.tex
\begin{table}
\centering
\begin{minipage}[t]{80mm}
\begin{center}
\begin{tabular}{@{\hspace{2pt}}l@{\hspace{10pt}}l@{\hspace{10pt}}l@{\hspace{10pt}}l@{\hspace{2pt}}}
\hline \noalign{\smallskip}
& Episode 1 & Episode 2 & Episode 3 \tabularnewline \noalign{\smallskip}
& (Old) & (Intermediate) & (Young) \tabularnewline
\hline
\hline \noalign{\smallskip}	
onset / yr ago & 4--$12 \times 10^9$ & 300--$400 \times 10^6$ & $10 \times 10^6$ \tabularnewline
$M_\star$ / $\textrm{M}_\odot$ & $\sim\,$$10^9$ & $\sim\,$$50\times10^6$ & $\sim\,$$3\times10^6$ \tabularnewline
SFR / $\textrm{M}_\odot\,\rm{yr}^{-1}$ & 0.25--0.10 & 0.14 & 0.28 \tabularnewline \hline
\end{tabular}

\end{center}
\vspace{-1mm}
\caption{The best-fitting three-episode star formation history of NGC~4449 assuming Kroupa IMF. The assumed star formation activity is continuous, and the onsets of episode 2 and~3 coincide with the end of the preceding episodes.\label{tab:mocassin_SFH}}
\end{minipage}
\end{table}

%% file: tables/table.MOCASSIN.lines.tex
\begin{table}
\centering
\begin{minipage}[t]{80mm}
\begin{center}
\begin{tabular}{@{\hspace{3pt}}l@{\hspace{5pt}}ccc@{\hspace{7pt}}c@{\hspace{3pt}}}

\hline
Line & Predicted & Observed & Ratio & References \tabularnewline
& $[I_{\mathrm{H}\beta}=1]$ & $[I_{\mathrm{H}\beta}=1]$ & & \tabularnewline \hline \hline

 [O~{\sc ii}] $\lambda$3727		& 3.198 & 3.891 & 0.822 & (1) \tabularnewline

 [Ne~{\sc iii}] $\lambda$3868		& 0.173 & 0.189 & 0.916 & (1)  \tabularnewline

 H$\delta$ $\lambda$4101		& 0.260 & 0.279 & 0.932 & (1)  \tabularnewline

 H$\gamma$ $\lambda$4340		& 0.469 & 0.492 & 0.953 & (1)  \tabularnewline

 H$\beta$ $\lambda$4861			& 1.000 & 1.000 & 1.000 & (1)  \tabularnewline

 [O~{\sc iii}] $\lambda$4959		& 0.699 & 0.689 & 1.015 & (1)  \tabularnewline

 [O~{\sc iii}] $\lambda$5007		& 2.086 & 2.069 & 1.008 & (1)  \tabularnewline

 He~{\sc i} $\lambda$5876		& 0.079 & 0.079 & 1.000 & (1)  \tabularnewline

 [S~{\sc iii}] $\lambda$6312		& 0.008 & 0.019 & 0.423 & (1)  \tabularnewline

 [N~{\sc ii}] $\lambda$6548		& 0.102 & 0.115 & 0.892 & (1)  \tabularnewline

 H$\alpha$ $\lambda$6563		& 2.922 & 2.866 & 1.020 & (1)  \tabularnewline

 [N~{\sc ii}] $\lambda$6584		& 0.313 & 0.338 & 0.927 & (1)  \tabularnewline

 He~{\sc i} $\lambda$6678		& 0.022 & 0.028 & 0.798 & (1)  \tabularnewline

 [S~{\sc ii}] $\lambda$6716		& 1.117 & 0.476 & 2.347 & (1)  \tabularnewline

 [S~{\sc ii}] $\lambda$6731		& 0.784 & 0.334 & 2.349 & (1)  \tabularnewline

 [S~{\sc iv}] 10.5 $\mu$m		& 0.264 & 0.219 & 1.205 & (2)  \tabularnewline

 [Ne~{\sc iii}] 15.6 $\mu$m		& 0.228 & 0.387 & 0.588 & (2)  \tabularnewline

 [O~{\sc i}] 63 $\mu$m			& 0.012 & 0.413 & 0.030 & (2)  \tabularnewline

 [O~{\sc iii}] 88 $\mu$m		& 0.654 & 0.647 & 1.011 & (2)  \tabularnewline 

 [C~{\sc ii}] 158 $\mu$m		& 1.112 & 1.307 & 0.851 & (2)  \tabularnewline \hline

\end{tabular}

\caption{Predicted and observed line strengths relative to H$\beta$ and the predicted-to-observed ratio for the best-fitting MOCASSIN model of NGC~4449. References: (1) \citet{Kobulnicky1999}; (2) \citetalias{Karczewski2013_OBSERVATIONS}. \label{tab:MOCASSIN_lines}}
\end{center}
\end{minipage}
\end{table}

%% file: figures/figure.mocassin.ionisation.tex
\begin{figure}
\centering
\begin{minipage}[t]{80mm}
\includegraphics[width=80mm]{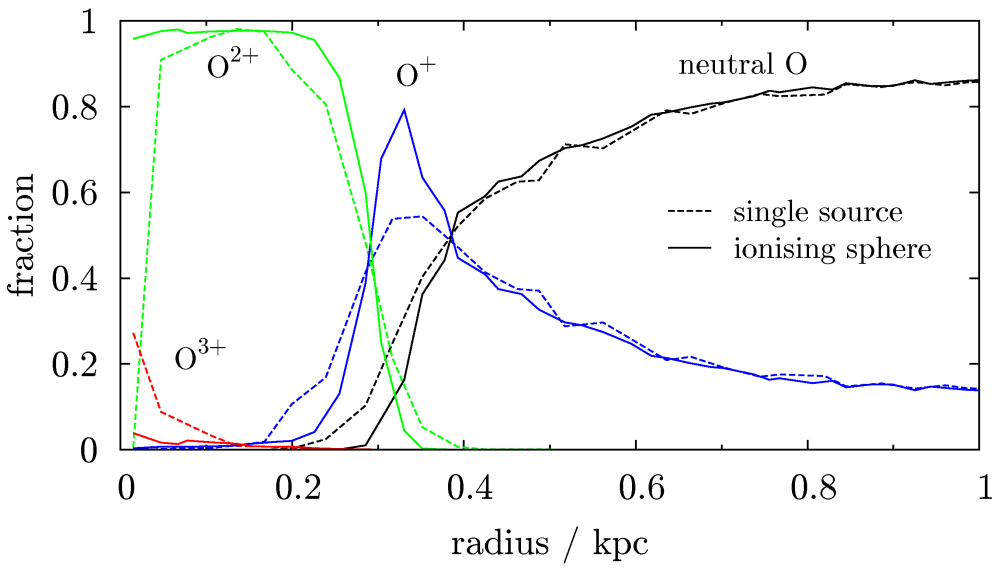}
\vspace{-5mm}
\caption{The ionization structure of oxygen as a function of radius
for the inner 1~kpc in the best-fitting MOCASSIN model of NGC~4449 (dashed lines) and 
the corresponding structure for the single ionizing source replaced
with 100 ionizing sources distributed uniformly
within a sphere of $r=0.2\;\textrm{kpc}$ (solid lines).\label{fig:n4449_mocassin_ionisation}}
\end{minipage}
\end{figure}

%% file: figures/figure.Swift.UVOT.tex
\begin{figure}
\centering
\begin{minipage}[t]{80mm}
\includegraphics[width=80mm]{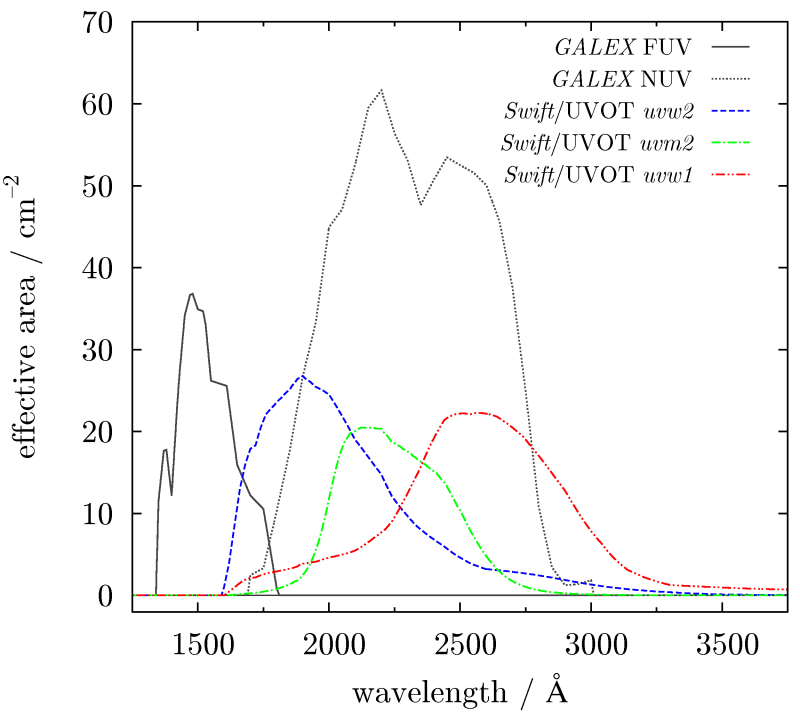}
\caption{Broad-band effective passbands for UV observations with \GALEX\ and \Swift.\label{fig:Swift_UVOT}}
\end{minipage}
\end{figure}

%% file: figures/figure.Swift.uvw2.tex
\begin{figure}
\centering
\begin{minipage}[t]{80mm}
\begin{center}
\includegraphics[width=70mm]{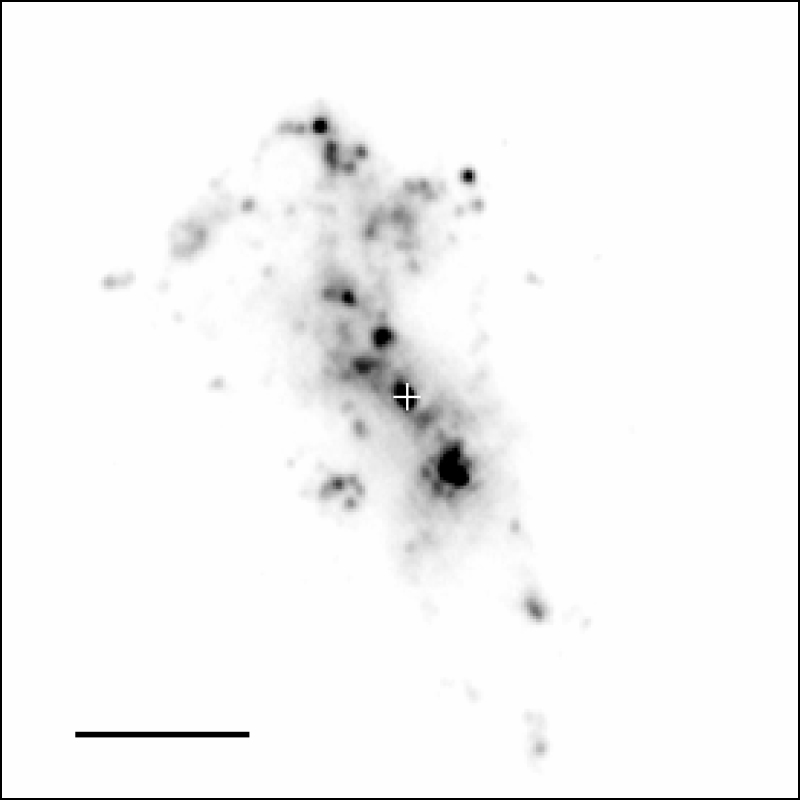}
\end{center}
\caption{Regions affected by coincidence loss in the {\it uvw2}-band image of NGC~4449. White indicates no coincidence loss. North is up, east is to the left. The bar is 1\arcmin\ in length and the cross denotes the optical centre of the galaxy.\label{fig:Swift_uvw2}}
\end{minipage}
\end{figure}

%% file: figures/figure.Swift.uvw2.regions.tex
\begin{figure}
\centering
\begin{minipage}[t]{80mm}
\begin{center}
\includegraphics[width=70mm]{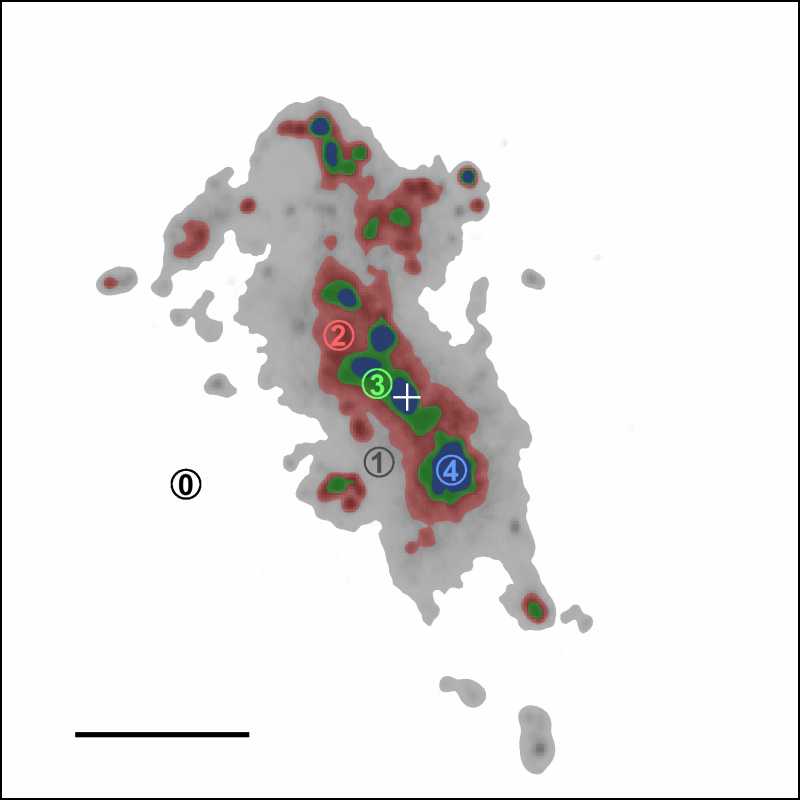}
\end{center}
\caption{The isophotal setup for the {\it uvw2}-band image of NGC~4449. Four isophotal regions numbered 1--4 are shown in grey, red, green and blue, respectively. Region~0, enclosed by an ellipse, and a background aperture defined as an elliptical annulus, both extend beyond the image and are not shown for clarity. The test regions are indicated by 5$^{\prime\prime}$ circles. North is up, east is to the left. The bar is 1\arcmin\ in length and the cross denotes the optical centre of the galaxy.\label{fig:Swift_uvw2_regions}}
\end{minipage}
\end{figure}

%% file: tables/table.appendix.uvw2.tex
\begin{table}
\centering
\begin{minipage}[t]{80mm}
\begin{center}
\begin{tabular}{cccccc}
\hline
$n$ & $A_{\mathrm{i},n}$ & $f_{\mathrm{test},n}$ & $C_{\mathrm{i},n}$ & $C_{\mathrm{src},n}$\tabularnewline
 & $[\mathrm{arcsec^{2}}]$ & & $[\mathrm{s}^{-1}]$ & $[\mathrm{s}^{-1}]$\tabularnewline
\hline
\hline

0 & 130000 & 1.003 & 525.6$\,\pm\,$1.4 & 409$\,\pm\,$21\tabularnewline
1 & 8580 & 1.034 & 470.2$\,\pm\,$1.1 & 479$\,\pm\,$5\tabularnewline
2 & 2870 & 1.086 & 396.8$\,\pm\,$0.9 & 428$\,\pm\,$3\tabularnewline
3 & 747 & 1.134 & 181.5$\,\pm\,$0.5 & 205$\,\pm\,$1\tabularnewline
4 & 426 & 1.494 & 206.8$\,\pm\,$0.2 & 309$\,\pm\,$1\tabularnewline
\hline
\end{tabular}

\end{center}
\vspace{-1mm}

\caption{Detailed photometric measurements for the five isophotal regions in NGC~4449
in band \textit{uvw2} shown in Fig.~\ref{fig:Swift_uvw2_regions}. $C_{\mathrm{bkg}} = 9.08 \pm 0.03 \times 10^{-4}$ s$^{-1}$~arcsec$^{-2}$. See Section~\ref{appendix:isophotal} for column definitions.
\label{tab:swift_photometry_uvw2}}
\end{minipage}
\end{table}


%% file: tables/table.appendix.photometry.tex
\begin{table}
\centering
\begin{minipage}[t]{80mm}
\begin{center}
\begin{tabular}{@{\hspace{1pt}}c@{\hspace{8pt}}c@{\hspace{8pt}}c@{\hspace{8pt}}c|c@{\hspace{1pt}}}
\hline \noalign{\smallskip}
 & \multicolumn{3}{c}{This work} & `as-is' \smallskip \tabularnewline
\cline{2-4} \noalign{\smallskip}
Band & $f_{\mathrm{max}}$ & $C_{\mathrm{src}}$ & $F_{\mathrm{src}}$ & $F_{\mathrm{src}}$ \tabularnewline
 & & $[\mathrm{s}^{-1}]$ & $[\mathrm{mJy}]$ & $[\mathrm{mJy}]$\tabularnewline
\hline
\hline
\emph{uvw2}/1991\r{A} & 1.49 & 1830$\,\pm\,$120 & 175$\,\pm\,$12 & 172$\,\pm\,$12 \tabularnewline
\emph{uvm2}/2221\r{A} & 1.32 & 1140$\,\pm\,$80 & 189$\,\pm\,$14 & 202$\,\pm\,$14 \tabularnewline
\emph{uvw1}/2486\r{A} & 1.43 & 1870$\,\pm\,$120 & 202$\,\pm\,$15 & 207$\,\pm\,$14 \tabularnewline
\emph{u}/3442\r{A} & 1.51 & 3480$\,\pm\,$300 & 249$\,\pm\,$22 & 468$\,\pm\,$40 \tabularnewline
\emph{b}/4321\r{A} & 1.40 & 4670$\,\pm\,$400 & 463$\,\pm\,$41 & 1200$\,\pm\,$100 \tabularnewline
\emph{v}/5410\r{A} & 1.18 & 2270$\,\pm\,$210 & 611$\,\pm\,$57 & 1250$\,\pm\,$100 \tabularnewline
\hline
\end{tabular}\end{center}
\vspace{-1mm}

\caption{Final count rates $C_{\mathrm{src}}$ and fluxes $F_{\mathrm{src}}$
in six UVOT bands for NGC~4449. $f_{\mathrm{max}}$ gives
the maximum correction factor used for each band. For comparison, given on the
right are the fluxes in the `as-is' approach.
All fluxes have been corrected for foreground extinction.\label{tab:swift_photometry_summary}}
\end{minipage}
\end{table}

%% file: figures/figure.Swift.SED.tex
\begin{figure}
\centering
\begin{minipage}[t]{80mm}
\includegraphics[width=80mm]{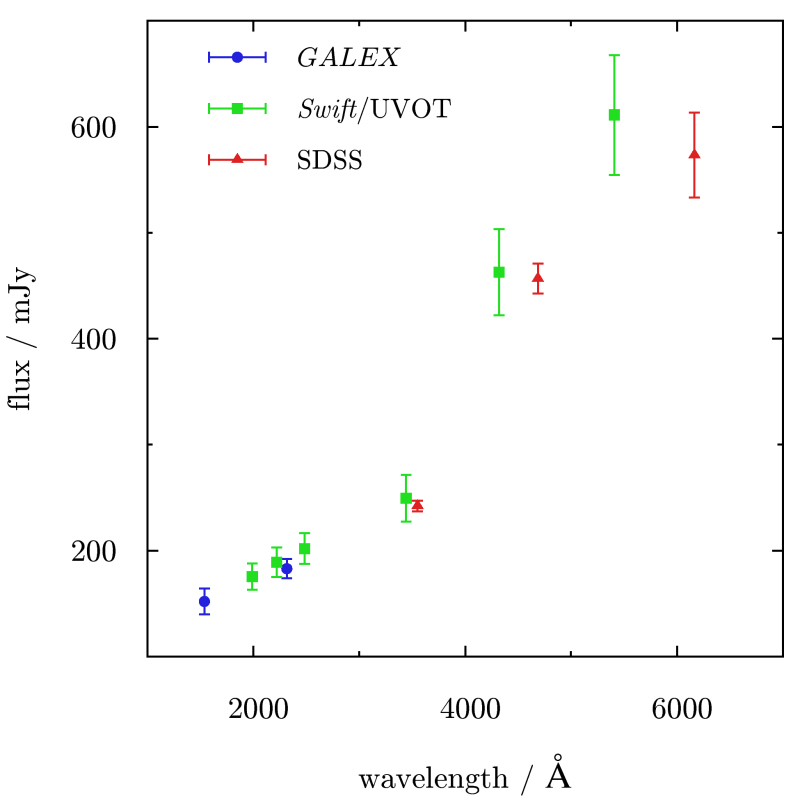}
\vspace{-4mm}
\caption{Global photometry of NGC~4449 with \GALEX, \Swift/UVOT and SDSS using data from Table~\ref{tab:photometry_table}.\label{fig:Swift_SED}}
\end{minipage}
\end{figure}